\documentclass[
preprintnumbers,
preprint,
a4paper,
showpacs, 
amsmath, 
amssymb, 
floatfix,
nofootinbib,
]{revtex4}

\usepackage[dvips]{graphicx}

\begin{document}

\title{Higher-dimensional numerical relativity:\\
Formulation and code tests}

\author{Hirotaka Yoshino}

\affiliation{Department of Physics, University of Alberta, 
Edmonton, Alberta, Canada T6G 2G7}

\author{Masaru Shibata}

\affiliation{Yukawa Institute for Theoretical Physics, 
Kyoto University, Kyoto, 606-8502, Japan}

\preprint{Alberta-Thy-10-09}

\date{July 16, 2009}


%
%
\begin{abstract}

We derive a formalism of numerical relativity for higher-dimensional
spacetimes and develop numerical codes for simulating a wide variety of
five-dimensional (5D) spacetimes for the first time.  First, the
Baumgarte-Shapiro-Shibata-Nakamura formalism is extended for arbitrary
spacetime dimensions $D \ge 4$, and then, the so-called cartoon method,
which was originally proposed as a robust method for simulating
axisymmetric 4D spacetimes, is described for 5D
spacetimes of several types of symmetries. Implementing 5D numerical
relativity codes with the cartoon methods, we perform test simulations
by evolving a 5D Schwarzschild spacetime and a 5D spacetime composed
of a gravitational-wave packet of small amplitude. The numerical
simulations are stably performed for a sufficiently long time, as done
in the 4D case, and the obtained numerical results agree well
with the analytic solutions: The numerical solutions are shown to
converge at the correct order. We also confirm that a longterm accurate
evolution of the 5D Schwarzschild spacetime is feasible using the
so-called puncture approach.  In addition, we derive the
Landau-Lifshitz pseudotensor in arbitrary dimensions, and show
that it gives a robust tool for computing the energy flux of 
gravitational waves.  The
formulations and methods developed in this paper provide a powerful
tool for studying nonlinear dynamics of higher-dimensional gravity.
\end{abstract}

\maketitle

%
%
\section{Introduction}

Clarifying the nature of higher-dimensional gravity has become an
important issue, since the braneworld scenarios were proposed
\cite{ADD98,RS99}.  If the space in which we live is a three-dimensional (3D) 
brane in extra spatial dimensions that are large or warped, 
the Planck
energy may be of $O({\rm TeV})$ and quantum gravity phenomena may
emerge in high-energy particle colliders such as the LHC. If this scenario
is correct, mini black holes may be produced at the LHC
\cite{BF99,DL01,GT02}, and this fact motivated a lot of theoretical
works in the past decade (see \cite{reviews} for a recent review). 
Understanding the AdS/CFT correspondence is
also an interesting issue in the higher-dimensional gravity.

To study nonlinear dynamics of spacetimes, numerical relativity is
probably the unique approach. In the past decade, numerical relativity
for four-dimensional (4D) spacetimes was significantly developed. 
Now, it is feasible to perform a longterm simulation for
merger of binary black holes or for high-velocity collision of two
black holes, which is one of the strongest gravitational
phenomena in nature (see
Refs.~\cite{P05,CLMZ06,BCCKM06,D06,HSL06,CORNELL} for pioneer works of
binary black hole merger). It is natural to expect that the
formulation and numerical techniques developed for 4D cases can be
extended to the higher-dimensional cases.  

There are also a few pioneer works in the five-dimensional (5D)
numerical relativity performed in the past decade
\cite{Choptuik:2003qd,Garfinkle:2004em}. However, the purpose of these
works was to study a specific issue, i.e., the Gregory-Laflamme
instability of a black string. Thus, the formulation and numerical
method in these works are applicable only for this particular issue,
and thus, developing a general formulation and codes in the
higher-dimensional numerical relativity is still an issue. 
Furthermore, there obviously remain a lot of interesting issues to be
explored in this field, as partially listed in the following.

The first issue is black hole formation in high-energy particle
collisions. If a black hole is formed at the LHC,
it will emit the Hawking radiation and may be detected.
To predict the rate of mini black hole production and
its detectability, it is necessary to know
the cross section for the black hole production $\sigma_{\rm BH}$ and
the resulting mass and angular momentum of the formed black hole.  
A partial answer was given in Refs.~\cite{YN03,YR05} (see also
\cite{EG02}) by numerically solving the apparent horizon at an instant
of the collision of Aichelburg-Sexl particles \cite{AS71} in higher
dimensions. Because the apparent horizon formation implies the
formation of the event horizon assuming the cosmic censorship 
(e.g., see \cite{Wald}), the cross section of the apparent
horizon formation $\sigma_{\rm AH}$ gives the lower bound of
$\sigma_{\rm BH}$.  However, the precise value of $\sigma_{\rm BH}$
itself is necessary for exactly predicting the phenomena in the
particle collider.  In the 4D case, high-velocity collisions of two
relativistic objects have been studied by full numerical relativity,
via a model of high-velocity collision of two black holes
\cite{Sperhake:2008ga,Shibata:2008rq,NEW}.  In particular,
Ref.~\cite{Shibata:2008rq} (see also Ref.~\cite{NEW} for the
refinement of the work of \cite{Shibata:2008rq}) first studied the
collisions with nonzero impact parameters and clarified that
$\sigma_{\rm BH}$ is approximately twice as large as $\sigma_{\rm AH}$
found in Ref.~\cite{YR05}.  They also showed that resulting mass and
angular momentum should be significantly smaller than the initial
values of the system because of a huge amount of gravitational
radiation. However, these studies are nothing but a prelude of the
study of high-velocity collisions in higher-dimensional spacetimes,
which is really required.

The second issue is on the stability of higher-dimensional rotating
black holes (Myers-Perry black hole) \cite{MP86}.  Although there are
works on the stability of those 
black holes by separating variables for the metric
perturbation in the linearized Einstein equation,
the analysis can be applied for limited situations
(see, e.g., \cite{Murata:2008yx} for special rotation parameters
and \cite{Oota:2008uj} for a tensor-mode perturbation)
and the problem has not been entirely investigated. 
Hence, multidimensional numerical
analyses are required.  Clarifying the stability of the
higher-dimensional rotating black holes with a single rotation
parameter, $a$, is important, because such a black hole would be the
outcome of particle collisions in the TeV gravity scenarios unless the
formed black hole is unstable. However, a recent numerical analysis
of the linearized Einsteins equation suggests that
black holes of high values of $a$ are unstable
at least for spacetime dimensions $D\ge 7$ \cite{Dias:2009iu} 
(see also \cite{Emparan:2003sy}): 
The production rate of mini black holes may be
smaller than what we naively expect from the analysis of the apparent
horizon \cite{YR05}.  To elucidate the stability and subsequent
evolution after the onset of instability, numerical relativity will
play a crucial role.

The third issue is on the evolution of a black hole on a
Randall-Sundrum (RS) brane.  So far, no analytic solution of a 5D static
black hole localized on the RS brane has been found.  The recent numerical
work \cite{Yoshino:2008rx} (see also \cite{Kudoh:2003xz}) indicates
the nonexistence of such solutions. If this is the case, any black
hole produced on the RS brane cannot relax to a stationary state but evolve
in time. Clarifying the fate of such black holes is an interesting
issue. Furthermore, if the AdS/CFT correspondence holds for the
RS models, a 5D classical black hole on the RS brane is dual to
a 4D black hole with quantum fields \cite{T02,EFK02}, and thus we
could obtain an indication for the Hawking radiation including the
back-reaction effects.

Motivated by these issues, we have developed numerical relativity
codes for simulating 5D spacetimes as the first step.  The purposes of
this paper are the following three.  The first purpose is to describe
a numerical relativity formulation in higher-dimensional spacetimes.
We adopt the Baumgarte-Shapiro-Shibata-Nakamura (BSSN) formalism
\cite{Shibata:1995we,Baumgarte:1998te} and write down its
higher-dimensional version.

The second purpose is to describe the cartoon method
\cite{Alcubierre:1999ab,Shibata2000} for 5D spacetimes of several
types of symmetries.  The cartoon method was originally proposed for
simulating 4D spacetimes of axial symmetry using the Cartesian
coordinates. This method has been demonstrated to be quite robust 
for accurately and stably simulating not only vacuum spacetimes 
but also rotating stars and rotating stellar core collapses (e.g.,
Ref.~\cite{Shibata2003}). The essence of this method is that we do not
have to use curvilinear coordinates that have coordinate singularities. In
most higher-dimensional problems, the spacetime should have
symmetries, e.g., among the extra-dimensional directions. For
such problems, it will be better not to adopt the curvilinear
coordinates but to adopt the Cartesian coordinates for an accurate 
and stable simulation, as we have learned in the 4D simulations. 
In the higher-dimensional issues that we listed above, 
several types of symmetries may be imposed. For example, in  
the off-axis collision of two black holes, the axes perpendicular to 
the orbital plane should be equivalent. In this paper, we 
particularly focus
on 4D spaces (i.e., 5D spacetimes) with $U(1)$ symmetry,
$U(1)\times U(1)$ symmetry, and $SO(3)$ symmetry.

The third purpose is to report our new codes for simulating 5D
spacetimes, which are implemented using the BSSN formalism and cartoon
methods.  For demonstrating that the codes work well, we perform
simulations for test problems for which analytical solutions are
known.  Specifically, evolutions for a 5D Schwarzschild spacetime and
for a 5D spacetime composed of a gravitational-wave packet of small
amplitude are chosen for the tests.  In the former case, we first solve
the 5D Schwarzschild spacetime by our codes in the geodesic slice and
show that the numerical results agree with the analytic solution
derived in this paper.  We also evolve the spacetime by the puncture
approach \cite{CLMZ06} with the dynamical slices and $\Gamma$-driver
conditions \cite{Alcubierre:2002kk}, and show that the longterm
evolution of a black hole spacetime is feasible as in the 4D case.  In
the second test, we compare the numerical results of a 
gravitational-wave packet with the semianalytic solution for 
linearized Einstein equations given in Appendix~A, 
and show that they agree well. In addition, we study a method for
estimating the energy flux carried by gravitational waves. The
Newman-Penrose formalism is widely used for extracting gravitational
waves in the 4D numerical relativity. Unfortunately, such formalism
has not yet been developed in higher dimensions. Here, we propose the
higher-dimensional Landau-Lifshitz pseudotensor \cite{LL75} for
calculating the energy flux carried by gravitational waves and
demonstrate that it correctly gives the amount of radiated energy.

This paper is organized as follows.  In Sec. II, we derive the BSSN
formalism in higher dimensions.  In Sec. III, we describe the cartoon
methods in 5D spacetimes of the three types of symmetries listed above.  In
Sec. IV, we present the numerical results of test simulations for 5D
numerical relativity and show that they agree with the analytic
solutions.  We also show that the energy extraction by the
Landau-Lifshitz pseudotensor works well.  Section V is devoted to a
summary.  In Appendix A, we summarize the equations and analytic
solutions for the linearized 5D Einstein equations 
of $U(1)\times U(1)$ symmetry or $SO(3)$ symmetry.  The 
Landau-Lifshitz pseudotensor in higher dimensions is given in
Appendix B.  Throughout the paper, we use the units of $c=1$ where $c$ 
is the speed of light. 

%
%
\section{BSSN formulation in higher dimensions}

In this section, we describe the BSSN formalism \cite{Shibata:1995we,
  Baumgarte:1998te} for higher-dimensional spacetimes. After reviewing
the Arnowitt-Deser-Misner (ADM) formulation for $D$ dimensions in
Sec.~IIA, the $D$-dimensional BSSN formalism is derived in Sec.~IIB.

%
%
\subsection{ADM formulation}

Suppose ${\cal M}$ be a $D$-dimensional spacetime with a metric
$g_{ab}$.  Consider a sequence of $(D-1)$-dimensional spacelike
hypersurfaces $\Sigma_t(\gamma_{ab}, K_{ab})$ foliated by a time
coordinate $t$ in ${\cal M}$. Here, $\gamma_{ab}$ is the induced
metric of $\Sigma_{t}$ defined by $\gamma_{ab}:=g_{ab}+n_an_b$ with
the future-directed unit normal $n_a$ to $\Sigma_t$.  $K_{ab}$ is the
extrinsic curvature $K_{ab}:=-(1/2)\mathcal{L}_n\gamma_{ab}$, where
$\mathcal{L}_n$ is the Lie derivative with respect to $n^a$.  The
coordinate basis $t^a$ of the time coordinate $t$ is decomposed as
$t^a=\alpha n^a+\beta^a$, where $\alpha$ is the lapse function and
$\beta^a$ is the shift vector.

The $D$-dimensional Einstein equation ${}^{(D)}G_{ab}=8\pi G_D T_{ab}$
is decomposed into constraint and evolution equations. 
Here, ${}^{(D)}G_{ab}$, $T_{ab}$, and $G_D$ are the $D$-dimensional 
Einstein tensor, the stress-energy tensor, and 
the gravitational constant, 
respectively. First, we define 
%
\begin{equation}
\rho:=T_{ab}n^an^b; 
\qquad j_a:=-T_{bc}n^b\gamma^{c}_{~a};
\qquad S_{ab}:=T_{cd}\gamma^{c}_{~a}\gamma^{d}_{~b}.
\end{equation}
%
The Hamiltonian constraint is derived from the Gauss equation to give 
%
\begin{equation}
R+K^2-K_{ab}K^{ab}=16\pi G_D \rho,
\label{Hamiltonian1}
\end{equation}
%
where $R$ is the Ricci scalar of the spacelike hypersurface $\Sigma_t$.
The momentum constraint is derived from the Codacci equation to give 
%
\begin{equation}
D_bK^{b}_{~a}-D_aK=8\pi G_D j_a,
\label{Momentum1}
\end{equation}
%
where $D_a$ denotes the covariant derivative with respect to $\gamma_{ab}$. 
The evolution equation of the induced metric $\gamma_{ab}$
is derived from the definition of the extrinsic curvature as
%
\begin{equation}
\mathcal{L}_t\gamma_{ab}
=-2\alpha K_{ab}+D_a\beta_b+D_b\beta_a,
\label{ADM-evolution1}
\end{equation}
%
and the evolution equation of the extrinsic curvature $K_{ab}$ is 
derived from the Ricci equation to give 
%
\begin{multline}
\mathcal{L}_t{K}_{ab}
=
-D_aD_b\alpha
+
\alpha\left( R_{ab}-2K_{ac}K^c_{~b}+K_{ab}K\right)
\\
+\beta^cD_cK_{ab}+K_{cb}D_a\beta^c
+K_{ca}D_b\beta^c
-8\pi G_D \alpha \left[
S_{ab} + \frac{\rho-S}{D-2}\gamma_{ab}
\right],
\label{ADM-evolution2}
\end{multline}
%
where $R_{ab}$ denotes the Ricci tensor with respect to $\gamma_{ab}$
and $S:=S^{c}_{~c}$.  The $D$-dimensional equations are formally different from 
the 4D equations only in the coefficient of the last term of 
Eq.~\eqref{ADM-evolution2}. In vacuum, the equations are independent of 
the value of $D$.

The above expressions are given in the covariant way. 
Here, we introduce coordinates $x^i$ that span the hypersurface
$\Sigma_t$, where $i,j=1,...,D-1$. In these coordinates, the line 
element is written by 
%
\begin{equation}
ds^2=-\alpha^2 dt^2
+\gamma_{ij}(dx^i+\beta^i dt)(dx^j+\beta^j dt),
\end{equation}
%
and the spatial components $\gamma^{ij}$ of $\gamma^{ab}$ are the
inverse of $\gamma_{ij}$.  The constraint and evolution equations in
the coordinate expressions are obtained just by replacing the indices
$a,~b$ to the spatial indices $i,~j$ and the Lie derivative
$\mathcal{L}_t$ to the coordinate derivative $\partial_t$ in 
Eqs.~\eqref{Hamiltonian1}, \eqref{Momentum1},
\eqref{ADM-evolution1}, and \eqref{ADM-evolution2}.

%
%
\subsection{BSSN formalism}

Now, we derive the BSSN formalism for higher-dimensional spacetimes. 
The basic idea
of the BSSN formalism is to increase the number of variables as well
as that of constraints in order to guarantee the stability 
in numerical computation (e.g., to kill constraint violation modes).
First, $\gamma_{ij}$ is conformally transformed as 
%
\begin{equation}
\tilde{\gamma}_{ij}=\chi\gamma_{ij},
\label{chi-def}
\end{equation}
%
where the conformal factor $\chi$ is chosen so that
the determinant of $\tilde{\gamma}_{ij}$ (denoted by $\tilde \gamma$) 
satisfies the condition 
%
\begin{equation}
\tilde{\gamma}=1.
\label{additional1}
\end{equation}
%
This is equivalent to setting $\chi=\gamma^{-1/(D-1)}$. 
We choose $\tilde \gamma_{ij}$ 
and $\chi$ as the fundamental variables. 

In the original BSSN formalism for the 4D spacetime, the conformal
factor $e^{-4\phi}$ was used rather than $\chi$. In the 4D puncture
formalism, $\chi$ or $W=\chi^{1/2}$ is often used \cite{CLMZ06}.  For
evolving the puncture black holes in five dimensions, $\chi$ turned
out to be a good choice. This is the reason that we choose $\chi$ as
one of the fundamental variables.

Next, the extrinsic curvature is decomposed into the trace part and
the trace-free part as
%
\begin{equation}
K_{ij}=A_{ij}+\frac{K}{D-1}\gamma_{ij},
\label{extrinsic-decomposition}
\end{equation}
%
where $K$ denotes the trace of $K_{ij}$ and $A_{ij}$ is the trace-free part. 
As in the 4D BSSN formalism, $K$ is chosen to be one of the fundamental variables. 
The trace-free part $A_{ij}$ is conformally transformed as
%
\begin{equation}
\tilde{A}_{ij}:=\chi A_{ij},
\label{def-tildeAij}
\end{equation}
%
and $\tilde A_{ij}$ is chosen to be one of the fundamental variables. 
Hereafter, the indices of $\tilde{A}_{ij}$ and $\tilde{A}^{ij}$ 
are raised and lowered by the conformally transformed metric
$\tilde{\gamma}^{ij}$ and $\tilde{\gamma}_{ij}$. 

In terms of the variables $\chi$, $K$, $\tilde{\gamma}_{ij}$,
and $\tilde{A}_{ij}$, the Hamiltonian constraint \eqref{Hamiltonian1} 
and the momentum constraint \eqref{Momentum1} are rewritten as
%
\begin{equation}
R+\frac{D-2}{D-1}K^2-\tilde{A}_{ij}\tilde{A}^{ij}=16\pi G_D \rho,
\label{Hamiltonian2}
\end{equation}
%
and
%
\begin{equation}
\partial_i\tilde{A}^{ij}
+\tilde{\Gamma}^{j}_{ik}\tilde{A}^{ik}
-\frac{D-2}{D-1}\tilde{\gamma}^{ij}K_{,i}
-\frac{(D-1)}{2}\frac{\chi_{,i}}{\chi}\tilde{A}^{ij}
=8\pi G_D \tilde{\gamma}^{ij}j_i,
\label{Momentum2}
\end{equation}
%
where $\tilde{\Gamma}^j_{ik}$ is the Christoffel symbol with respect
to $\tilde{\gamma}_{ij}$ and the comma ($,i$) denotes the derivative
by $x^i$.

The evolution equation of $\chi$ is derived from 
Eq.~\eqref{ADM-evolution1} with Eqs.~\eqref{chi-def} and \eqref{additional1}
to give
%
\begin{equation}
(\partial_t-\beta^i\partial_i)\chi
=\frac{2}{D-1}\chi\left(\alpha K-\partial_i\beta^i\right).
\label{evolution-chi}
\end{equation}
%
Multiplying $\gamma^{ab}$ to Eq.~\eqref{ADM-evolution2}
and rewriting it with Eqs.~\eqref{extrinsic-decomposition},
\eqref{def-tildeAij}, and \eqref{Hamiltonian2},
the evolution equation of $K$ is derived to give 
%
\begin{equation}
(\partial_t -\beta^i\partial_i)K
=-D_i D^i \alpha
+\alpha\left(\tilde{A}^{ij}\tilde{A}_{ij}+\frac{K^2}{D-1}\right)
+\frac{8\pi\alpha}{D-2}\left[(D-3)\rho+S\right].
\label{evolution-K}
\end{equation}
%
Rewriting Eq.~\eqref{ADM-evolution1} with Eqs.~\eqref{chi-def},
\eqref{extrinsic-decomposition}, \eqref{def-tildeAij},
and \eqref{evolution-chi}, the evolution equation of the conformal 
$(D-1)$-metric is derived as 
%
\begin{equation}
(\partial_t
-\beta^k\partial_k)\tilde{\gamma}_{ij}
=
-2\alpha\tilde{A}_{ij}
+\tilde{\gamma}_{ik}\partial_j\beta^k
+\tilde{\gamma}_{jk}\partial_i\beta^k
-\frac{2}{D-1}\partial_k\beta^k\tilde{\gamma}_{ij}.
\label{evolution-gammaij}
\end{equation}
%
The evolution equation of $\tilde{A}_{ij}$ is derived by substituting
Eq.~\eqref{extrinsic-decomposition} with Eqs.~\eqref{chi-def} and
\eqref{def-tildeAij} into Eq.~\eqref{ADM-evolution2} and using
Eqs.~\eqref{Hamiltonian2}, \eqref{evolution-chi}, \eqref{evolution-K},
and \eqref{evolution-gammaij}, to give 
%
\begin{multline}
(\partial_t-\beta^k\partial_k)\tilde{A}_{ij}
=\chi\left[-(D_iD_j\alpha)^{\rm TF}
+\alpha\left(R_{ij}^{\rm TF}-8\pi S_{ij}^{\rm TF}\right)\right]
\\
+\alpha\left(K\tilde{A}_{ij}-2\tilde{A}_{ik}\tilde{A}^k_j\right)
+\tilde{A}_{ik}\partial_j\beta^k
+\tilde{A}_{kj}\partial_i\beta^k
-\frac{2}{D-1}\partial_k\beta^k\tilde{A}_{ij},
\label{evolution-Aij}
\end{multline}
%
where TF denotes the trace-free part, e.g.
$R_{ij}^{\rm TF}=R_{ij}-R\gamma_{ij}/(D-1)$.

The Ricci tensor is decomposed into two parts as
%
\begin{equation}
R_{ij}=\tilde{R}_{ij}+R_{ij}^{(\chi)},
\end{equation}
%
where $\tilde{R}_{ij}$ is the Ricci tensor with respect to
$\tilde{\gamma}_{ij}$ and $R_{ij}^{(\chi)}$ is the contribution of the
conformal factor.  Here, $\tilde{R}_{ij}$ has the terms
$(1/2)\tilde{\gamma}^{kl}
(\tilde{\gamma}_{kj,il}+\tilde{\gamma}_{il,kj}
-\tilde{\gamma}_{kl,ij}-\tilde{\gamma}_{ij,kl})$ and the first three
terms could be the source of a numerical instability, as often found
in 4D numerical relativity.  For stable numerical integration, 
the following auxiliary variable is introduced \cite{Baumgarte:1998te}:
%
\begin{eqnarray}
\tilde{\Gamma}^i&:=&\gamma^{jk}\tilde{\Gamma}^{i}_{jk}
=-\tilde{\gamma}^{ik}_{~~,k}.
\label{additional3}
\end{eqnarray}
%
We note that another choice 
$F_i:= \delta^{jk} \partial_k \tilde \gamma_{ij}$ 
can be used as well \cite{Shibata:1995we}. It was  
found that the numerical results for the test simulations 
in this paper do not essentially depend on the choice. 

Using the variable $\tilde \Gamma^i$, $\tilde{R}_{ij}$ and $R_{ij}^{(\chi)}$
are rewritten as
%
\begin{multline}
\tilde{R}_{ij}=
-\frac12\tilde{\gamma}^{kl}\tilde{\gamma}_{ij,kl}
+\frac12\left(\tilde{\gamma}_{ki}\partial_j\tilde{\Gamma}^k+
\tilde{\gamma}_{kj}\partial_i\tilde{\Gamma}^k\right)
\\
-\frac12\biggl(
\tilde \gamma_{il,k}\tilde \gamma^{kl}_{~~,j}
+\tilde \gamma_{jl,k}\tilde \gamma^{kl}_{~~,i}
-\tilde \Gamma^l\tilde \gamma_{ij,l}
\biggr)
-\tilde \Gamma^l_{ik} \tilde \Gamma^k_{jl}, 
\label{bar-Ricci}
\end{multline}
%
and 
%
\begin{multline}
R_{ij}^{(\chi)}=
\frac{(D-3)}{2\chi}
\left(\chi_{,ij}-\tilde{\Gamma}^{k}_{ij}\chi_{,k}\right)
-\frac{(D-3)}{4}\frac{\chi_{,i}\chi_{,j}}{\chi^2}
\\
+\tilde{\gamma}_{ij}\tilde{\gamma}^{kl}
\left[\frac{\chi_{,kl}}{2\chi}-(D-1)\frac{\chi_{,k}\chi_{,l}}{4\chi^2}\right]
-\frac12\tilde{\gamma}_{ij}\frac{\chi_{,m}}{\chi}\tilde{\Gamma}^{m},
\label{Ricci-chi}
\end{multline}
%
where $\tilde \gamma=1$ is used in deriving these equations. 
As in the 4D case, the second derivatives of $\tilde{\gamma}_{ij}$
explicitly appear only in the first term of Eq.~\eqref{bar-Ricci}.

Since $\tilde{\Gamma}^i$ is one of the dynamical variables 
in the BSSN formalism,
its evolution equation has to be derived.
Substituting Eq.~\eqref{evolution-gammaij} 
into $\partial_t\tilde{\Gamma}^i=\partial_j
(\tilde{\gamma}^{ik}\tilde{\gamma}^{jl}\tilde{\gamma}_{kl,t})$
and eliminating $\tilde{A}^{ij}_{~~,j}$ with 
Eq.~\eqref{Momentum2}, we obtain
%
\begin{multline}
(\partial_t-\beta^j\partial_j)\tilde{\Gamma}^i
=-2\tilde{A}^{ij}\partial_j\alpha
+2\alpha\left[
\tilde{\Gamma}^{i}_{jk}\tilde{A}^{jk}
-\frac{D-2}{D-1}\tilde{\gamma}^{ij}K_{,j}
-8\pi\tilde{\gamma}^{ij}j_j
-\frac{(D-1)}{2}\frac{\chi_{,j}}{\chi}\tilde{A}^{ij}
\right]
\\
-\tilde{\Gamma}^j\partial_j\beta^i
+\frac{2}{D-1}
\tilde{\Gamma}^i\partial_j\beta^j
+\frac{D-3}{D-1}\tilde{\gamma}^{ik}\beta^j_{,jk}
+\tilde{\gamma}^{jk}\beta^i_{,jk}.
\label{evolution-Gammai}
\end{multline}
%

In summary, the variables to be evolved are $\chi$, $K$,
$\tilde{\gamma}_{ij}$, $\tilde{A}_{ij}$ and $\tilde{\Gamma}^i$ (or
$F_i$), and their evolution equations are Eqs.~\eqref{evolution-chi},
\eqref{evolution-K}, \eqref{evolution-gammaij}, \eqref{evolution-Aij},
and \eqref{evolution-Gammai}, respectively.  The conditions
\eqref{additional1}, ${\rm tr}(\tilde A_{ij})=0$, and
\eqref{additional3} are regarded as the new constraints which arise 
because the dynamical variables are increased.

As shown above, the BSSN formalism for higher dimensions 
essentially has the same form as that for the 4D case, except that some
coefficients are changed. Because of the change in the coefficients, the
behavior of the solutions near the black hole and in the wave zone is 
changed. However, this change does not significantly affect the
stability and accuracy in numerical computations
at least for evolutions
of the 5D Schwarzschild spacetime and a 5D spacetime of
small-amplitude gravitational waves as shown in
Sec. IV.

%
%
\section{Cartoon method}

In this section, we describe the cartoon method for 5D spacetimes. The
cartoon method was originally proposed as a prescription for stable
numerical simulations of axisymmetric 4D spacetimes. The essence in
this method is not to use curvilinear coordinates but to use the Cartesian
coordinates \cite{Alcubierre:1999ab}. We briefly review this (say, the
case ``$x=y,\ z$'') in Sec.~IIIA.  Next, we extend this method to 5D
spacetimes with symmetries. In higher-dimensional spacetimes, there 
are various types of possible symmetries.  Here, we consider 4D spaces
(i.e., 5D spacetimes) with $U(1)$ symmetry (the case ``$x,\ y,\
z=w$''), $U(1)\times U(1)$ symmetry (the case ``$x=y,\ z=w$''), and
$SO(3)$ symmetry (the case ``$x=y=z,\ w$''). The cartoon methods for
these three cases are described in Secs.~IIIB, IIIC, and IIID,
respectively.

%
%
\subsection{3D axisymmetric space}

For 3D axisymmetric spacelike hypersurfaces in a 4D spacetime, the 3D
Cartesian coordinates $(x,y,z)$ 
can be introduced so that 
the vector
$\partial_{\varphi}:=x\partial_y-y\partial_x$ 
becomes the Killing vector. 
In other words, each spacelike hypersurface has $U(1)$ symmetry around the
$z$ axis. We refer to this case as $x=y,\ z$ in short. 

One natural coordinate choice for this space is the cylindrical
coordinates $(\rho, \varphi, z)$.  If these coordinates are adopted,
the problem reduces to a 2D problem (i.e., all quantities 
depend only on $\rho$ and $z$). However, in these coordinates, 
the symmetry axis $\rho=0$ is the coordinate singularity. 
On this coordinate singularity, one has to change the manner of 
finite differencing because there is no point of $\rho <0$. 
This sometimes (not always) causes a numerical instability, 
which is known as the finite discretization instability.
Although it might be possible to stabilize numerical computations 
by appropriately modifying the finite-differencing method,  
there is the case that the prescription is not simple or has not been 
found without numerical viscosity, e.g., issues for which 
a longterm simulation of rotating objects is necessary. 

One can avoid this problem by using the Cartesian coordinates because
they have no coordinate singularities. The shortcoming in the
Cartesian coordinates is that $U(1)$ symmetry does not explicitly
appear in equations, and thus, we have to solve 3D equations.
Suppose that the initial data are given on the
$(x,z)$  plane (i.e., $\varphi=0$). In the case that the cylindrical
coordinates are adopted, the subsequent evolution of the system is
feasible with this data. However, in the Cartesian
coordinates, one cannot calculate the next step only with this data,
because the equations include $y$ derivatives of functions to be
solved.

However, we do not have to prepare the data for all values of $y$,
if the cartoon method is used. In this method, a few grid
points in the neighborhood of the $(x,z)$  plane are prepared.  The
number of necessary grid points depends on the order of numerical accuracy
required in the finite differencing (see the last paragraph of this 
subsection). Then, the data at a grid point $(x, y\neq 0,z)$
are generated using the data at a point $(\rho, 0, z)$
[i.e., on the $(x,z)$  plane],
where $\rho=\sqrt{x^2+y^2}$,
by use of the symmetry. 
Because the grid point is not located at
the point $(\rho, 0, z)$ in general, 
the data at this point
are determined by an interpolation. The method of the interpolation
depends on the required order of accuracy (see the last paragraph
of this subsection).
Once the data at
the grid points $y\neq 0$ are known, 
$y$ derivatives at $y=0$ are calculated 
and the data on the $(x,z)$ plane are evolved toward the next time
step.

The symmetric relations are derived by the fact that
the Lie derivative of functions with respect to the Killing vector
becomes zero. For a scalar function $\Psi(x,y,z)$,
the symmetric relation is 
%
\begin{equation}
\Psi(x,y,z)=\Psi(\rho,0,z).
\label{henkan-axi-scalar}
\end{equation} 
%
In order to derive the symmetric relation of a contravariant 
vector function $T^i$, it is convenient to
consider the coordinate transformation from the $(\rho,\varphi)$ 
coordinates to the Cartesian coordinates.  After expressing 
$T^x(x,y,z)$ and $T^y(x,y,z)$ in terms of $T^{\rho}(\rho,z)$ and 
$T^{\varphi}(\rho,z)$, the latter two can be replaced by the relations 
on the $(x,z)$ plane, $T^x(\rho, 0, z)=T^\rho(\rho, z)$ and $T^y(\rho,
0, z)=\rho T^\varphi(\rho, z)$.  This yields 
%
\begin{eqnarray}
T^x(x,y,z) &=& 
(x/\rho) T^x(\rho, 0, z)-(y/\rho) T^y(\rho, 0, z), 
\label{henkan-axi-vector1}
\\
T^y(x,y,z) &=& 
(y/\rho) T^x(\rho, 0, z)+(x/\rho) T^y(\rho, 0, z).
\label{henkan-axi-vector2}
\end{eqnarray}
%
The relation between $T^z(x,y,z)$ and $T^z(\rho,0,z)$ is the same as that
for a scalar, described in Eq.~\eqref{henkan-axi-scalar}.
A covariant vector $T_i(x,y,z)$
has the same symmetric relation as that of
$T^i(x,y,z)$.

In a similar manner, the symmetric relation of 
a symmetric covariant tensor function
$S_{ij}=S_{(ij)}$ is obtained. 
$S_{zz}(x,y,z)$ has the same relation as that for a
scalar, Eq.~\eqref{henkan-axi-scalar}, and $S_{zx}$ and $S_{zy}$
have the same relations as $x$ and $y$ components of a vector function,
Eqs.~\eqref{henkan-axi-vector1} and \eqref{henkan-axi-vector2}.  For
the other components, the following relations are derived: 
%
\begin{eqnarray}
S_{xx}(x,y,z) &=& 
(x/\rho)^2 S_{xx}(\rho, 0, z)+(y/\rho)^2 S_{yy}(\rho, 0, z)
-(2xy/\rho^2)S_{xy}(\rho,0,z), 
\label{henkan-axi-tensor1}
\\
S_{yy}(x,y,z) &=& 
(y/\rho)^2 S_{xx}(\rho, 0, z)+(x/\rho)^2 S_{yy}(\rho, 0, z)
+(2xy/\rho^2)S_{xy}(\rho,0,z),
\label{henkan-axi-tensor2} 
\\
S_{xy}(x,y,z) &=& 
(xy/\rho^2) \left[S_{xx}(\rho, 0, z)-S_{yy}(\rho, 0, z)\right]
+[(x^2-y^2)/\rho^2] S_{xy}(\rho,0,z).
\label{henkan-axi-tensor3}
\end{eqnarray}
%
Again, a contravariant symmetric tensor $S^{ij}$
has the same symmetric relation
as that of $S_{ij}$.

Using the above relations, the data for $y \not=0$ are generated using
the data in the $(x,z)$  plane, and thus, the derivatives with respect
to $y$ can be calculated. The required grid number is 5 for the
fourth-order finite differencing (i.e., the data at $y=\pm \Delta y$ and
$\pm 2\Delta y$ have to be determined, where $\Delta y$ is the grid 
spacing), and 3 for the second-order one. For obtaining the 
values at a point $(\rho, z)$
on the $(x,z)$  plane, interpolation is necessary. To keep the 
fourth-order accuracy, we have to use at least fourth-order accurate 
interpolation (e.g., fourth-order Lagrangian interpolation). 

%
%
\subsection{4D space with $U(1)$ symmetry}

In the following, we describe three cartoon methods in 5D spacetimes
(4D spaces) of three types of symmetries, denoting the Cartesian
coordinates by $(x,y,z,w)$ and assuming that the 4D space is
topologically identical to the 4D Euclidean space.

First, we consider a 4D space of $U(1)$ symmetry whose corresponding 
Killing vector is $\partial_\psi=z\partial_w-w\partial_z$ 
(i.e., $\tan\psi=w/z$).
An example of a system of this symmetry is an off-axis collision of
two black holes. Suppose that the centers of the two black holes are
located in the $(x, y)$  plane.  In this case, the directions
orthogonal to the $(x,y)$  plane (i.e., the direction of $z$ and $w$
axes) are equivalent, and thus, the space has the $U(1)$ symmetry. We
refer to this symmetric space as $x,\ y,\ z=w$ in short.

Such a spacetime can be simulated as a 3+1 problem using the cartoon
method in a similar prescription to that in the 3D axisymmetric space.
We first prepare the grid points in the $(x,y,z)$ plane and a few
neighboring grid points in the $w$ direction.  Then, the data at a point
$(x, y, z, w\neq 0)$ are generated by the data at a point $(x,y,\rho, 0)$ with
symmetric relations, where $\rho=\sqrt{z^2+w^2}$. The 
symmetric relations are
essentially same as those in the 3D axisymmetric case: 
It is sufficient to 
replace the indices $(x, y)$ in
Eqs.~\eqref{henkan-axi-vector1}--\eqref{henkan-axi-tensor3}
to $(w, z)$.
As for the other components, 
$T^x$, $T^y$, $S_{xx}$, $S_{yy}$, and $S_{xy}$ behave like scalar
functions, and 
$(S_{xz}, S_{xw})$ and $(S_{yz}, S_{yw})$ behave like $z$
and $w$ components of vector functions $(T^z, T^w)$.

%
%
\subsection{4D space with $U(1)\times U(1)$ symmetry}

Next, we describe the cartoon method in a 4D space of $U(1)\times
U(1)$ symmetry, where two Killing vector fields,
$\partial_\varphi=x\partial_y-y\partial_x$ and
$\partial_\psi=z\partial_w-w\partial_z$, are present. 
An example of a spacetime of this
symmetry is a 5D rotating black hole spacetime with two rotation
parameters \cite{MP86}.  In this spacetime, the black hole is rotating
with respect to the $(x,y)$- and $(z,w)$ planes simultaneously. 
Of course, a 5D rotating black
hole with one rotation parameter is also the case for this
symmetry.  We refer to such a case as ``$x=y$, $z=w$'' in short.

The spacetime of this symmetry can be simulated as a 2+1 problem in
the cartoon method. We prepare grid points on the $(x,z)$ plane and a
few neighboring grid points in both $y$ and $w$ directions. In this
symmetry, two cartoons are necessary: 
The first cartoon to generate the data in the $y$- direction,
and the second cartoon to generate the data in the $w$- direction.  
The symmetric relations
for each cartoon are essentially the same as those in the previous two
subsections.

In numerical simulation, we have the data of points $(x,0,z,0)$ at
each time step. Then, we apply the first cartoon to generate the data
for grid points $(x,y,z,0)$.  After that, we apply the second cartoon
to generate the data for grid points $(x,y,z,w)$. Then, all the
necessary derivatives with respect to $y$ and $w$ can be taken and
the data can be evolved to the next time step. This method may be
called the {\it double} cartoon method.  As we demonstrate in
Sec. IV, the double cartoon method works well as the single 
cartoon method.

%
%
\subsection{4D space with $SO(3)$ symmetry}

Finally, we consider a space of a different type of symmetry, $SO(3)$
symmetry, in which the three Killing vectors,
$\boldsymbol{\xi}_1:=y\partial_z-z\partial_y$,
$\boldsymbol{\xi}_2:=z\partial_x-x\partial_z$, and
$\boldsymbol{\xi}_3:=x\partial_y-y\partial_x$ are present.  
In other words, each
hypersurface of $w={\rm const.}$ is spherically symmetric. An example
for a spacetime of this symmetry is a head-on collision of two black
holes moving along the $w$ axis, because the other directions $x$,
$y$, and $z$ are equivalent 
if the black holes are not rotating.  
We refer to this case as $x=y=z,\ w$ in short. 

This spacetime can be simulated as a 2+1 problem in the cartoon
method.  We prepare grid points in the $(x,w)$ plane and a few
neighboring grid points in both $y$ and $z$ directions. Using the data
on the $(x,w)$ plane, the data at points $(x,y,z,w)$ can be calculated
by the $SO(3)$ symmetric relations using the data at the point
$(r,0,0,w)$ where $r=\sqrt{x^2+y^2+z^2}$.
The data at the point $(r,0,0,w)$ should be determined by an  
interpolation as before.

In this case, the symmetry relations are different from those in the
previous cases. For scalar functions, it is trivial as
%
\begin{equation}
\Psi(x,y,z,w)=\Psi(r,0,0,w).
\label{henkan-spheri-scalar}
\end{equation}
%
In order to derive the symmetry relations for vector and symmetric
tensor functions, we have to know the $SO(3)$-symmetric forms
of a vector and a symmetric tensor, 
which can be found by the conditions
$\mathcal{L}_{\boldsymbol{\xi}_n}T^i=0$ and
$\mathcal{L}_{\boldsymbol{\xi}_n}S_{ij}=0$.  For this purpose, 
we first describe their components in the spherical-polar coordinates
$(r,\theta,\varphi,w)$ introduced by $x=r\sin\theta\cos\varphi$,
$y=r\sin\theta\sin\varphi$, and $z=r\cos\theta$. Then,
in the $SO(3)$ symmetry, they are written as 
%
\begin{equation}
T^i(r,\theta,\varphi,w)=(T^r(r,w),\ 0,\ 0,\ T^w(r,w)),
\end{equation}
and 
%
%
\begin{equation}
S_{ij}(r,\theta,\varphi,w)=\left(
\begin{array}{cccc}
S_{rr}(r,w) & 0 & 0 & S_{rw}(r,w) \\
* & S_{\theta\theta}(r,w) & 0 & 0 \\
* & * & S_{\theta\theta}(r,w)\sin^2\theta & 0 \\
* & * & * & S_{ww}(r,w)
\end{array}
\right).
\end{equation}
%
Now we transform these expressions to the Cartesian coordinates
and use the relations on the $(x,w)$ plane
to give
%
\begin{eqnarray}
T^x(x,y,z,w)&=&(x/r)T^x(r,0,0,w),\\
T^y(x,y,z,w)&=&(y/r)T^x(r,0,0,w),\\
T^z(x,y,z,w)&=&(z/r)T^x(r,0,0,w),
\end{eqnarray}
%
for a vector function and
%
\begin{eqnarray}
S_{xx}(x,y,z,w)&=&(x^2/r^2) S_{xx}(r,0,0,w)
+(1-x^2/r^2)S_{yy}(r,0,0,w),
\\
S_{yy}(x,y,z,w)&=&(y^2/r^2) S_{xx}(r,0,0,w)
+(1-y^2/r^2)S_{yy}(r,0,0,w),
\\
S_{zz}(x,y,z,w)&=&(z^2/r^2) S_{xx}(r,0,0,w)
+(1-z^2/r^2)S_{yy}(r,0,0,w),
\end{eqnarray}
%
%
\begin{eqnarray}
S_{yz}(x,y,z,w)&=&
(yz/r^2)
[S_{xx}-S_{yy}](r,0,0,w),
\\
S_{zx}(x,y,z,w)&=&
(zx/r^2)
[S_{xx}-S_{yy}](r,0,0,w),
\\
S_{xy}(x,y,z,w)&=&
(xy/r^2)
[S_{xx}
-S_{yy}](r,0,0,w),
\end{eqnarray}
%
for a tensor function. Note that $T^w$ and $S_{ww}$ satisfy the
symmetry relation of a scalar function, and $(S_{wx}, S_{wy}, S_{wz})$
satisfy that of $(x,y,z)$ components of a vector function.

Using these symmetry relations, one can calculate the data at grid
points in the neighborhood of the $(x,w)$ plane (i.e., $y, z\neq 0$), 
and thus the evolution can be 
performed as a 2+1 problem.  Note that by eliminating the $w$
direction, the above symmetry relations can be used also for
simulating a 3D spherically symmetric space in a 4D spacetime. 

%
%
\section{Code tests}

In the previous two sections, we have described necessary ingredients
for higher-dimensional numerical relativity, i.e., the BSSN formalism and
the cartoon method.  Based on these, we have implemented several codes
for simulating 5D spacetimes in the following manner. As often done
in the 4D numerical relativity (e.g., Ref.~\cite{BHBH}), we adopt the
centered fourth-order finite differencing in the space directions,
except the advection terms such as $\beta^k\partial_k \tilde
\gamma_{ij}$ for which the fourth-order upwind finite differencing is
adopted.  The time evolution is carried out using the fourth-order
Runge-Kutta method, where the Courant number is adopted to be 
$0.5$. Vertex-centered grids are employed for all the space 
directions.  In the present codes, we do not implement adaptive mesh
refinement (AMR) algorithm. We plan to combine our codes with our AMR
code (SACRA code) in the future \cite{SACRA}.

So far, we have developed the 3D codes for spacetimes with $U(1)$
symmetry ($x,\ y,\ z=w$), the 2D codes for spacetimes
with $U(1)\times U(1)$ symmetry 
($x=y,\ z=w$) and with $SO(3)$ symmetry ($x=y=z,\ w$).  The 
authors of this paper have independently developed the codes, and
checked that the numerical results for test simulations derived by the two
codes agree. In addition, Yoshino has made a 1D code for spacetimes with
$SO(4)$ symmetry (``$x=y=z=w$'' in short). 
In the following, we present the results
by Yoshino's code, for which the uniform grid with the grid spacing $\Delta
x$ is always adopted for all the space directions.

In order to prove the validity of our codes, we consider that at least
the following two test simulations have to be successfully carried out as
in the 4D numerical relativity.  One is the evolution of the 5D
Schwarzschild black hole, and the other is the evolution of a
spacetime composed of gravitational waves of small amplitude.  Since
their metrics are analytically given, they can be used in the
benchmark tests for calibrating the codes.  The results of the test
simulations are reported in Secs. IVA and IVB, respectively.  In
addition, we show in Sec. IVB that the energy flux can be correctly
calculated for linear gravitational waves using the Landau-Lifshitz
pseudotensor.

%
%
\subsection{5D Schwarzschild spacetime}

First, we analytically derive the 5D Schwarzschild metric in the 
geodesic slices and then compare numerical results with it.  Next, we
demonstrate that a long-term evolution of the 5D Schwarzschild
spacetime is feasible in the so-called puncture approach, as in the 4D
case (e.g., Refs.~\cite{CLMZ06,BHBH,Hannam:2006vv,Hannam:2006xw}).

%
%
\subsubsection{Geodesic slices}

The well-known metric of a 5D Schwarzschild spacetime is 
%
\begin{equation}
ds^2=-f(r)dt^2+\frac{dr^2}{f(r)}+r^2d\Omega_3^2,
\qquad
f(r)=1-\frac{r_h^2}{r^2},
\label{5D-Schwarzschild}
\end{equation}
%
where $d\Omega_3^2$ is the line element of a 3D unit sphere
and $r_h$ is the Schwarzschild radius 
%
\begin{equation}
r_h=\sqrt{\frac{8 G_5 M}{3 \pi}}.
\end{equation}
%
Here, we consider the Gaussian normal coordinates 
starting from the $t=0$ hypersurface, 
which is analogous to the Novikov coordinates 
in the 4D Schwarzschild spacetime \cite{Novikov,MTW73}. 
Let us introduce a geodesic congruence of 
test particles that are initially at rest.
Each geodesic labels the radial coordinate. Denoting $r_0$
as the initial value of $r$ for each geodesic, we define $\bar r$ by
%
\begin{equation}
r_0=\bar r\left(1+\frac{r_h^2}{4{\bar r}^2}\right)
\label{isotropic}
\end{equation}
%
as the radial coordinate. At $t=0$, the coordinate $\bar r$ agrees 
with the so-called isotropic radial coordinate. Adopting 
the proper time $\tau$ for each geodesic as the time coordinate, 
the geodesic equations are solved to give 
%
\begin{equation}
r^2=r_0^2-\left({r_h}/{r_0}\right)^2\tau^2,
\label{geodesic-r}
\end{equation}
and 
%
%
\begin{equation}
t=\sqrt{f(r_0)}\tau
+\frac{r_h}{2}
\log\left|
\frac{\tau+(r_0^2/r_h)\sqrt{f(r_0)}}{\tau-(r_0^2/r_h)\sqrt{f(r_0)}}
\right|.
\label{geodesic-t}
\end{equation}
%
Substituting these equations into Eq.~\eqref{5D-Schwarzschild} and
transforming $r_0$ to $\bar r$ with Eq.~\eqref{isotropic}, we obtain
%
\begin{equation}
ds^2=-d\tau^2+\frac{\left[r_0^2+(r_h/r_0)^2\tau^2\right]^2}
{\left[r_0^2-(r_h/r_0)^2\tau^2\right]}
\frac{d{\bar r}^2}{{\bar r}^2}
+\left[r_0^2-\left({r_h}/{r_0}\right)^2\tau^2\right]d\Omega_{3}^2.
\label{geodesicslice}
\end{equation}
%
This line element shows that the $\bar r\bar r$ component of the
metric diverges at $\tau=r_0^2/r_h$. Curvature invariants indeed show that
the curvature singularity appears at this time. This implies that 
at the time $\tau=r_h$, the slice hits the singularity at 
$\bar r=r_h/2$. 

The derived line element \eqref{geodesicslice} shows the exact solution, 
and thus it can be used for test simulations. In this test, we 
perform a simulation with the gauge condition $\alpha=1$ and $\beta^i=0$, 
until the computation crashes approximately at the crash time $\tau_{\rm crash}=r_h$.

%
\begin{figure}[tb]
\centering
{
\includegraphics[width=0.45\textwidth]{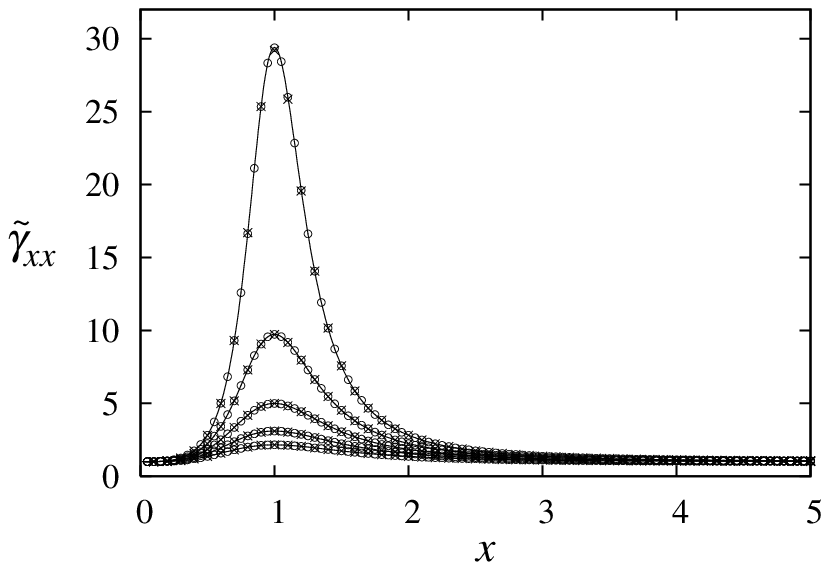}
\includegraphics[width=0.45\textwidth]{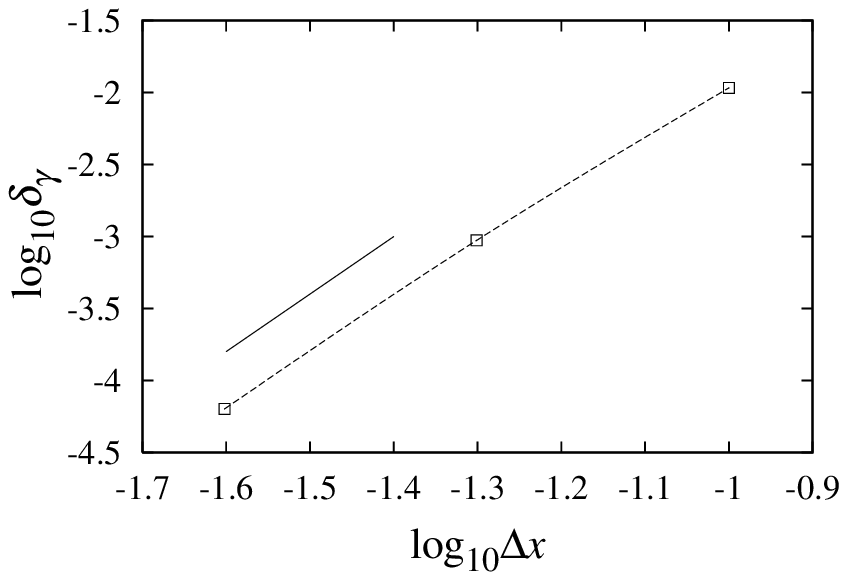}
}
\caption{Left-hand panel: Snapshots of $\tilde{\gamma}_{xx}$ along the
  $x$ axis for $\tau/r_h=0.5$, 0.6, 0.7, 0.8, and $0.9$. The unit of
  $x$ is $r_h/2$. The grid resolutions are $\Delta x=0.1$ ($\times$)
  and $0.05$ ($\odot$). The solid curves denote the analytic solutions, 
  $\tilde{\gamma}^{\rm (a)}_{xx}$.  Right-hand panel: The averaged
  error, $\delta_{\gamma}$, as a function of $\Delta x$. Here, the
  average is taken for the data in the range $0\le x\le 5$.  The upper
  short line segment shows the relation of the fourth-order
  convergence (i.e., a segment with the slope $4$).}
\label{gaussian_evolution_Gxx}
\end{figure}
%

The left-hand panel of Fig.~\ref{gaussian_evolution_Gxx} shows the
snapshots of $xx$ component of the conformal 4D metric
$\tilde{\gamma}_{xx}$ along the $x$ axis for various time slices as
$\tau/r_h=0.5$, 0.6, 0.7, 0.8, and 0.9.  For this plot, the grid
resolutions $\Delta x=0.1$ and $0.05$ are adopted.  Here, the units
of $x$ are $r_h/2$ (i.e., the event horizon is initially located at
$x=1$). We see that the values of $\tilde{\gamma}_{xx}$ rapidly
increase and blow up around $x=1$, and agree approximately with the
analytic solutions \eqref{geodesicslice} (solid curves).

The right-hand panel of Fig.~\ref{gaussian_evolution_Gxx} 
plots the averaged error as a function of the grid spacing $\Delta x$. 
Here, the averaged error is defined by
%
\begin{eqnarray}
\delta_{\gamma}=
\frac{1}{5} \int_0^5 dx |\tilde{\gamma}_{xx}-\tilde{\gamma}^{\rm (a)}_{xx}|,
\label{ave}
\end{eqnarray}
%
where $\tilde{\gamma}_{xx}$ and $\tilde{\gamma}^{\rm (a)}_{xx}$ are 
numerical and analytic solutions, respectively, 
and the integral is numerically performed using the data on the grid points.
This figure indicates 
that the numerical error approaches zero approximately at the fourth-order 
convergence. 

Here, the figures are plotted for the results 
obtained by the 1D code ($x=y=z=w$), but essentially
the same results are obtained by the 2D codes
($x=y$, $z=w$ and $x=y=z$, $w$)
and the 3D code ($x$, $y$, $z=w$).

%
\begin{figure}[tb]
\centering
{
\includegraphics[width=0.45\textwidth]{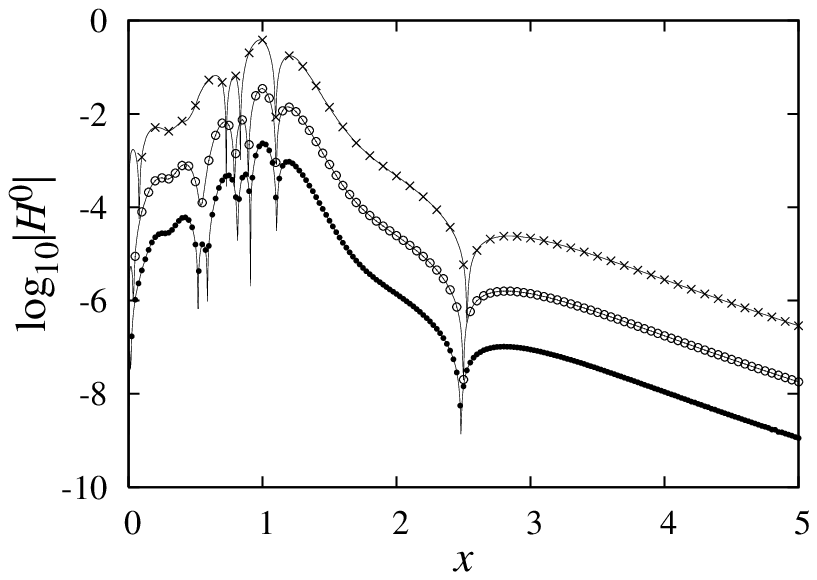}
\includegraphics[width=0.45\textwidth]{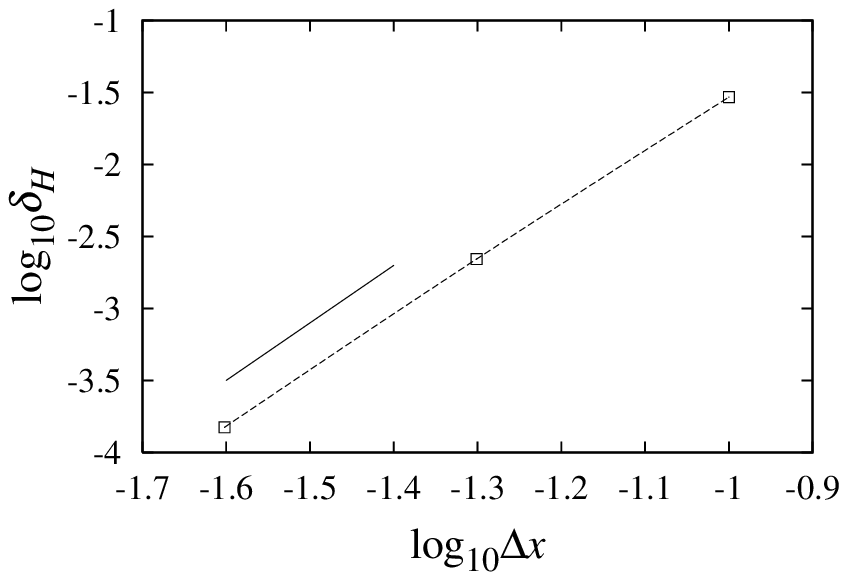}
}
\caption{Left-hand panel: The violation of the Hamiltonian constraint at
  the time $\tau/r_h=0.9$ for the grid resolutions $\Delta x=0.1$
  ($\times$), $0.05$ ($\odot$), and $0.025$ ($\bullet$).
  Although the violation grows as the 4D hypersurface approaches the
  singularity, it becomes smaller for a fixed value of $\tau$ as the resolution
  is increased.  Right-hand panel: The average, $\delta_{H}$, as a function
  of the grid spacing $\Delta x$. The upper short line segment shows the
  relation of the fourth-order convergence (i.e., a segment with the slope $4$).} 
\label{gaussian_H0_T0.9}
\end{figure}
%

The left-hand panel of 
Fig.~\ref{gaussian_H0_T0.9} plots the violation of 
the Hamiltonian constraint along the $x$ axis at 
$\tau/r_h=0.9$. Here, the violation is defined by 
%
\begin{equation}
H^0:=R-\frac{3}{4}K^2+\tilde{A}_{ij}\tilde{A}^{ij}. 
\label{def:H0}
\end{equation}
%
As the surface approaches the singularity, the value of the 
constraint violation grows rapidly. However, if we fix the time
$\tau$ and compare the results by the different grid resolutions 
$\Delta x=0.1$ ($\times$), $0.05$ ($\odot$), and $0.025$ ($\bullet$),
the clear convergence is seen. 

The right-hand panel of Fig.~\ref{gaussian_H0_T0.9} plots the 
averaged constraint violation $\delta_H$. Here, the average 
is defined in the same manner as Eq.~\eqref{ave}. 
This figure indicates that the error converges also at the 
fourth order approximately.

%
%
\subsubsection{Long-term evolution}

Next, we show that long-term evolution of a black hole is feasible
using the puncture approach as in the 4D case. In this test, the initial
condition is prepared in the isotropic coordinates, and then, the
evolution is carried out without excising the black hole interior.  As
the gauge conditions, we adopt the generalized version of the 
dynamical slicing condition \cite{Alcubierre:2002kk,Hannam:2006xw}
%
\begin{equation}
\partial_t \alpha = -\eta_\alpha \alpha K,
\label{modified-1+log}
\end{equation}
%
and the $\Gamma$-driver gauge condition \cite{Alcubierre:2002kk}
%
\begin{equation}
\partial_t\beta^i= \frac{(D-1)}{2(D-2)}v_{long}^2 B^i,
\qquad
\partial_tB^i=\partial_t\tilde{\Gamma}^i-\eta_\beta B^i. 
\end{equation}
%
Here, $v_{long}$ indicates the propagation speed of a gauge
mode and has to be chosen as $0<v_{long}\le 1$. We tried the choices
$v_{long}=1$ and $\sqrt{3}/2$, and found that the stable 
numerical evolution is possible in both cases. 
The choice $v_{long}=\sqrt{3}/2$ stabilizes the 
numerical evolution near the puncture a little more. 
$\eta_\alpha$ and $\eta_\beta$ are positive constants 
that can be arbitrarily chosen.
For $\eta_\alpha$, we chose several values between 1.2 and 2.0, and found that the 
stable and long-term simulation is feasible irrespective of 
the value of $\eta_\alpha$. For $\eta_\beta$, we choose $1/5r_h$. 


%
\begin{figure}[tb]
\centering
{
\includegraphics[width=0.45\textwidth]{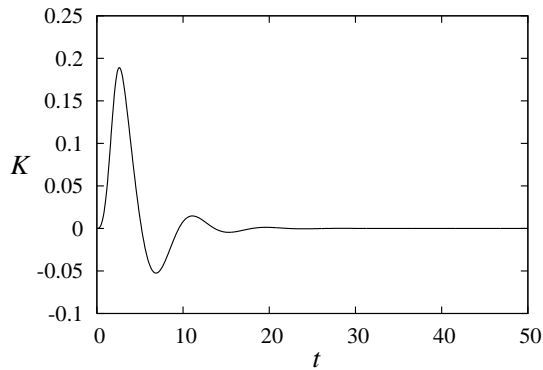}
}
\caption{The evolution of the trace of 
the extrinsic curvature $K$ at $x=r_h/2$ ($y=z=w=0$) 
in the puncture gauge. The unit of $t$ is $r_h/2$.
The value of $K$ asymptotes to zero.}
\label{5DSch-onepluslog-TRK-evolution}
\end{figure}
%

%
\begin{figure}[tb]
\centering
{
\includegraphics[width=0.45\textwidth]{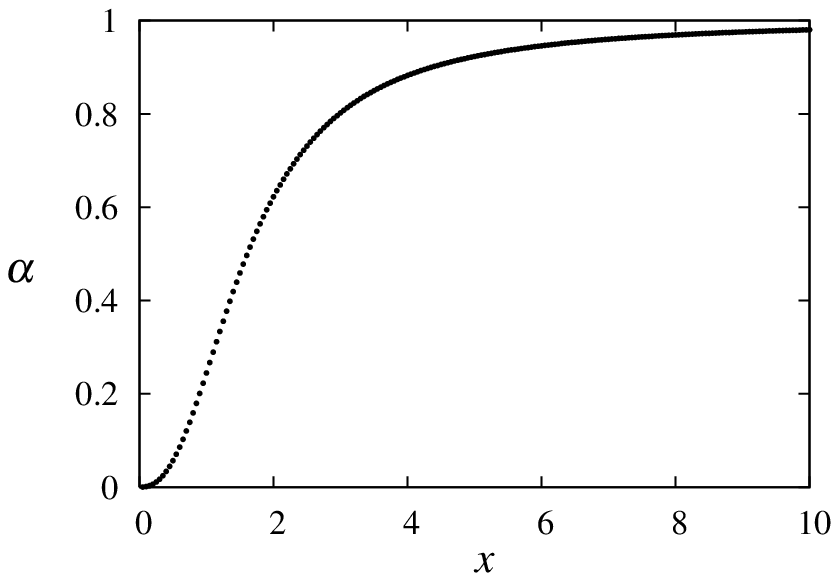}
\hspace{5mm}
\includegraphics[width=0.45\textwidth]{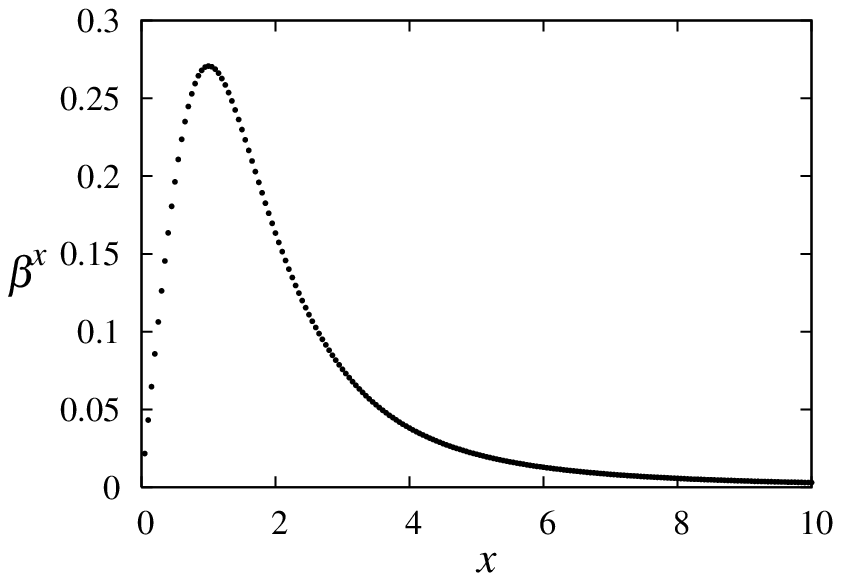}
}
\caption{The values of the lapse $\alpha$ and the shift vector $\beta^x$
along the $x$ axis at the time $t=50r_h$. Here, the unit of $x$
is $r_h/2$. }
\label{5DSch-onepluslog-dx0.05-T100}
\end{figure}
%

In the following, we show the results of the numerical evolution for
the case $v_{long}=1$ and $\eta_\alpha=2$. The initial condition of
the lapse and shift is chosen as $\alpha=\sqrt{\chi}$ and $\beta^i=0$.
Figure~\ref{5DSch-onepluslog-TRK-evolution} shows the evolution of $K$
at $x=r_h/2$ on the $x$ axis.  The unit of the length is $r_h/2$.  The
value of $K$ relaxes to zero after a few oscillations, and the slice
asymptotes to a maximal surface because of the property of the
dynamical slicing condition~\eqref{modified-1+log}.  We evolved this
spacetime up to $t=100r_h$, and the spacetime relaxes to a 
stationary state. Figure~\ref{5DSch-onepluslog-dx0.05-T100} shows the values
of $\alpha$ and $\beta^x$ along the $x$ axis at $t=50r_h$. By
this time, the variables approximately relax to stationary states. 
These results are quite similar to the evolution of a 4D
Schwarzschild spacetime (compare with Figs.~1 and 5 in Ref. 
\cite{Hannam:2006xw}).

%
\begin{figure}[tb]
\centering
{
\includegraphics[width=0.45\textwidth]{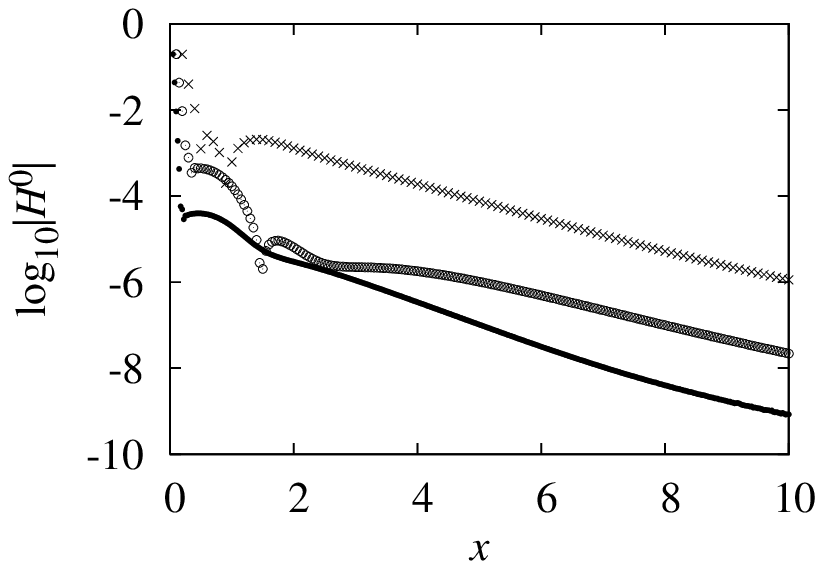}
\includegraphics[width=0.45\textwidth]{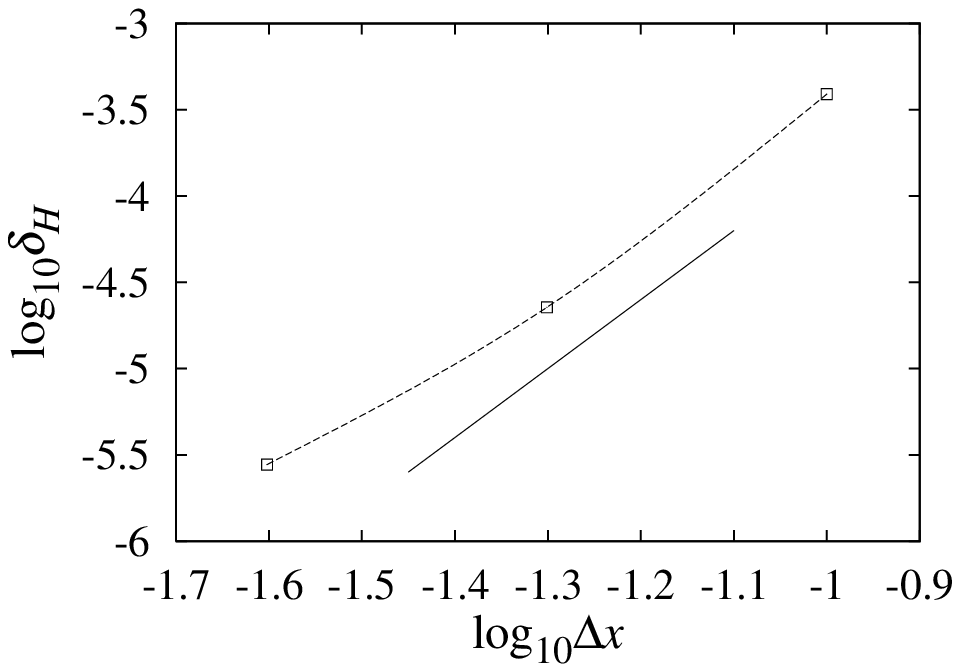}
}
\caption{Left-hand panel: The violation of the Hamiltonian constraint at
  the time $t=50r_h$ 
  for the grid resolutions $\Delta x=0.1$
  ($\times$), $0.05$ ($\odot$), and $0.025$ ($\bullet$).
  After the long-term evolution, the spatial pattern of $H^0$ depends 
  on the resolution, and the error generated at the puncture is $O(1)$.
  But the general tendency is that the violation becomes smaller 
  as the resolution is increased.  
  Right-hand panel: The average, $\delta_{H}$, in the range
  $0.5\le x\le 10$ as a function of the grid spacing $\Delta x$. 
  The lower short line segment shows the
  relation of the fourth-order convergence (i.e., a segment with the slope $4$).
  The convergence is worse than the fourth-order convergence because of
  the error generated at the puncture.} 
\label{longterm_H0}
\end{figure}
%

The left-hand panel of Fig.~\ref{longterm_H0}
plots the violation from the Hamiltonian constraint 
defined by Eq.~\eqref{def:H0} along the $x$ axis at 
the time $t=50r_h$ for the grid resolutions
$\Delta x=0.1$ ($\times$), $0.05$ ($\odot$), and $0.025$ ($\bullet$).
After the long-term evolution, the violation in the neighborhood
of the puncture $x=0$ grows large to become $O(1)$.
This is because the analyticity of the solution is violated
at the puncture. However, the error rapidly decreases as 
$x$ is increased, indicating the reliability of the numerical results. 
It is also found that 
the spatial patterns of $H^0$ depend on the 
resolution after the longterm evolution, $t\gg r_h$,
although initially they have similar shapes.

The right-hand panel of Fig.~\ref{longterm_H0} plots the 
averaged constraint violation $\delta_H$ in the range
$0.5\le x\le 10$ defined in the same manner as Eq.~\eqref{ave}. 
Because of the error generated at the puncture,
the value of $\delta_H$ does not show the fourth-order
convergence. Nevertheless, the violation rapidly decreases
as the grid resolution is increased.

The obtained stationary data are expected to agree with those of the
limit surface of the maximally sliced evolution. In the 4D case, the
limit surface of $K=0$ was analytically determined
\cite{Estabrook:1973ue,Baumgarte:2007ht} and also the asymptotic
solution in the numerical simulation agrees with it.  A simulation 
\cite{Hannam:2006xw} also demonstrates that the spacetime remains 
in a stationary state if the limit surface is adopted as the
initial condition.  The limit surface exists also in a higher-dimensional
Schwarzschild spacetime and it provides a useful benchmark for
calibrating codes for higher dimensions, as shown in Ref.~\cite{NASY09}. 

%
%
\subsection{Linear gravitational waves}

We turn our attention to a simulation for propagation of gravitational
waves of small amplitude.  Here, we focus only on gravitational waves
that preserve $U(1)\times U(1)$ symmetry ($x=y$, $z=w$) or $SO(3)$
symmetry ($x=y=z$, $w$).  In Appendix A, the linearized Einstein
equations of such symmetries and their special solutions for the
lowest multipole moment $l$ are described.  In this subsection, we
pick up a tensor-mode perturbation with $U(1)\times U(1)$ symmetry for
the test simulation.

%
\begin{figure}[tb]
\centering
{
\includegraphics[width=0.45\textwidth]{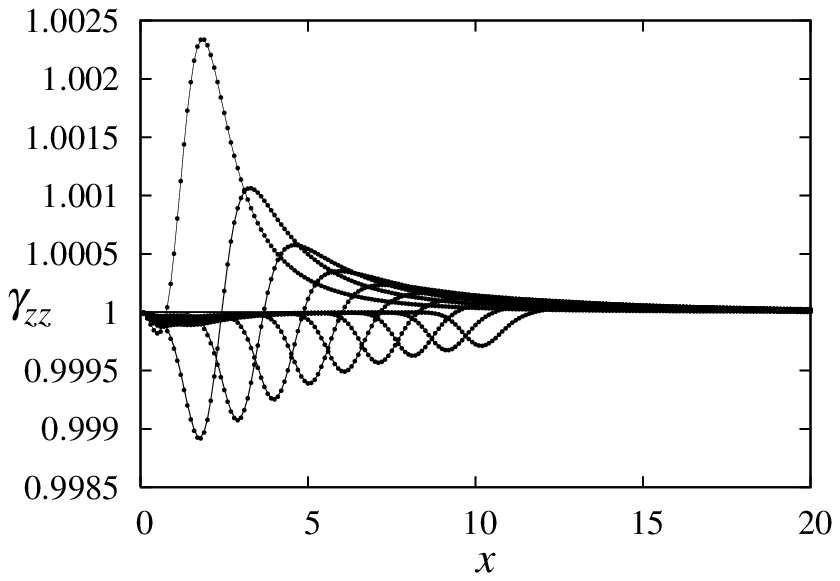}
\includegraphics[width=0.45\textwidth]{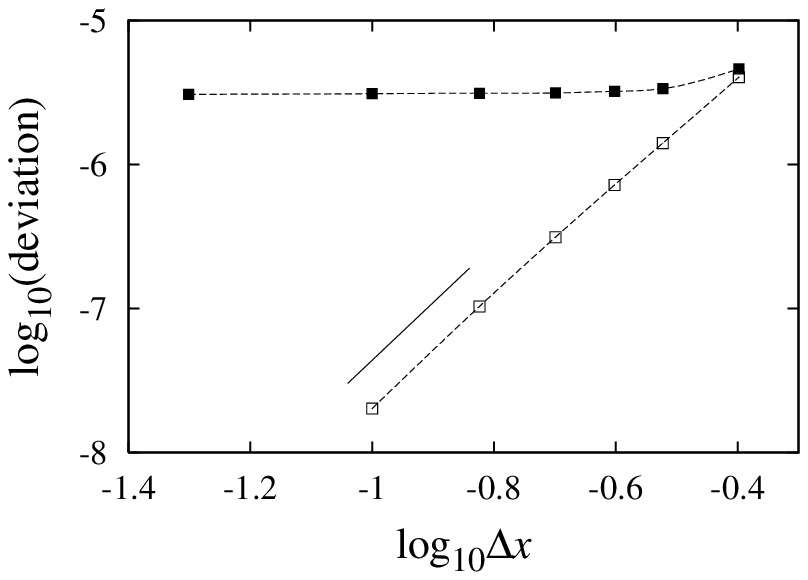}
}
\caption{Left-hand panel: Snapshots of $\gamma_{zz}$ along the $x$ axis for
  $t=1,2,...,10$ for propagation of a gravitational-wave packet. The
  dotted points ($\bullet$) and solid curves denote the numerical results and
  perturbative solutions, respectively.  Right-hand panel: The averaged
  deviation of $\gamma_{xx}$ from the analytic perturbative solution
  ($\blacksquare$) and that from the numerical data with the grid
  resolution $\Delta x=0.05$ ($\square$) as functions of $\Delta x$.
  Here, the data at time $t=3$ are used and the average is taken for
  the data in the range $0\le x\le 6$ and $0\le z\le 6$.  The upper
  short line segment shows the relation of the fourth-order
  convergence (i.e., a segment with the slope $4$).} 
\label{Tensor_Gzz_Xaxis}
\end{figure}
%

As the perturbative solution used for the test simulation, we 
adopt the spatial metric \eqref{metric-tensor-mode} with
Eq.~\eqref{harmonic-tensor} and the special solution for $h(t,r)$ 
given by Eq.~\eqref{special-solution-u} with the gauge condition
$\alpha=1$ and $\beta^i=0$.  Here, we set $A=1/6$ and
$B=0$ in Eq.~\eqref{harmonic-tensor}, and $A_0=0.015$ and $\omega_0=2$
in Eq.~\eqref{special-solution-u}.  In the simulation, we evolved
the initial data that correspond to the perturbative solution
under the same gauge condition $\alpha=1$ and $\beta^i=0$.  
The left-hand panel of
Fig.~\ref{Tensor_Gzz_Xaxis} compares the analytic solution for the
linearized Einstein equation (solid curves) and the numerical results
(dotted points) 
obtained by the 2D ``$x=y$, $z=w$'' code (where
the double cartoon method is used). The values of
$\gamma_{zz}$ are plotted along the $x$ axis for
$t=1,2,...,10$.  The two results agree well, indicating the validity of
our code.

The right-hand panel of Fig.~\ref{Tensor_Gzz_Xaxis} 
shows the deviation of the
numerical solution from the analytic solution of the perturbation as a
function of grid sizes $\Delta x$ (black squares, $\blacksquare$).
Here, we used the data of $\gamma_{xx}$ on the $(x,z)$ plane at $t=3$
and evaluated the deviation by taking the average of
$|\gamma_{xx}-\gamma_{xx}^{(a)}|$ in the region $0\le x\le 6$ and
$0\le z\le 6$. The deviation scarcely depend on $\Delta x$, and thus
it is not caused by the grid resolution.  The deviation primarily comes
from the fact that the perturbative solution ignores the
second- and higher-order quantities in $h_{ij}$, whereas the numerical
simulation is carried out by the fully nonlinear evolution equations. 
Indeed, the order of 
the difference $\sim 10^{-5}$ agrees with the magnitude of the 
nonlinear effect for our chosen wave amplitude. 

Squares ($\square$) in the same panel
show the difference
of the numerical data computed with the grid resolutions $\Delta x\ge 0.1$
from the one with the grid resolution $\Delta x=0.05$. The
difference decreases approximately at the fourth order, implying that
our numerical solution achieves the fourth-order convergence.

Using the analytic solution of the linearized Einstein equation, we
can also test extraction methods of gravitational-wave energy flux
from the numerical data.  The Newman-Penrose variable is now widely
used for extracting radiated energy of gravitational waves in the 4D
numerical relativity.  However, the formalism for the extraction based
on the Weyl scalar has not been developed in higher dimensions. Thus,
for the calculation of the energy flux, we adopt the Landau-Lifshitz
pseudotensor $t_{\rm LL}^{\mu\nu}$ \cite{LL75}, whose $D$-dimensional
form is given in Appendix B.  Because $t_{\rm LL}^{\mu\nu}$ is not a
tensor, it is a coordinate-dependent quantity in general. However, as
discussed in p.~85 of Ref.~\cite{Wald}, the total amount of gravitational
energy and the total radiated energy, Eqs.~\eqref{Etot} and
\eqref{Erad}, are shown to be the gauge-invariant quantities for
the linear gravitational waves of a 
perturbed flat spacetime up to second order
with respect to the metric perturbation. 

Let $r=r_{\rm obs}$ be the radius of the extraction of
gravitational-wave energy flux. The integrated energy flux $E_{\rm
  rad}$ through the surface $r=r_{\rm obs}$ is evaluated by 
%
\begin{equation}
E_{\rm  rad}(r_{\rm obs})=\int t_{\rm LL}^{0r} dS dt, 
\label{Erad-linear}
\end{equation}
%
where $dS$ is the area element of a hypersphere of $r=r_{\rm obs}$.
Here, the second-order expression of
the Landau-Lifshitz pseudotensor 
with respect to the perturbative quantities 
is used (see Eq.~\eqref{tLL-perturbation} and subsequent explanation).
We evaluate $E_{\rm rad}(r_{\rm obs})$ both for the perturbative solution
and for the full numerical solution, and compare the two results.

%
\begin{figure}[tb]
\centering
{
\includegraphics[width=0.6\textwidth]{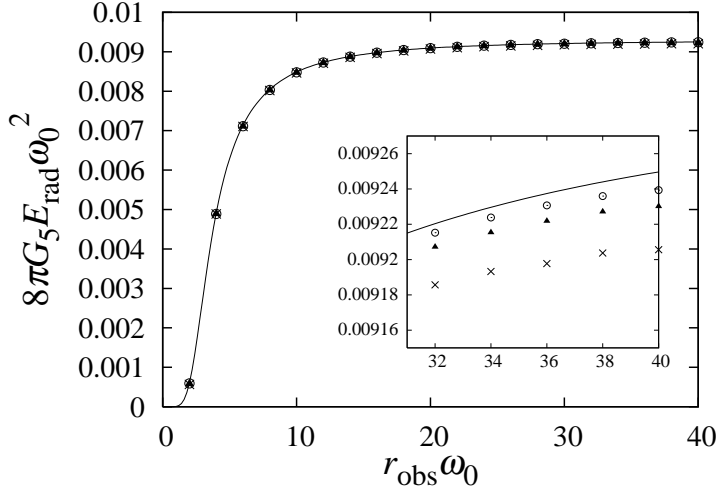}
}
\caption{The integrated energy flux $8\pi G_5 E_{\rm rad}\omega_0^2$ 
  calculated
  by the Landau-Lifshitz pseudotensor as a function of the
  extraction radii $r_{\rm obs} \omega_0$ for the analytic solution (solid
  curve) and for the numerical results computed with the grid resolutions
  $\Delta x=0.1$ ($\odot$), $0.15$ ($\blacktriangle$), and $0.2$ ($\times$).
  The inset shows the enlarged figure for the region 
  $31\le r_{\rm obs}\omega_0\le 40$.}
\label{RE_linear_gaussian_onepluslog}
\end{figure}
%

The value of $E_{\rm rad}(r_{\rm obs})$
for the perturbative solution is evaluated
semianalytically by proceeding the integrations 
of Eq.~\eqref{Erad-linear} numerically.  
Figure~\ref{RE_linear_gaussian_onepluslog} shows the value of 
$8\pi G_5 E_{\rm rad} \omega_0^2$ as a function of 
$r_{\rm obs} \omega_0$ 
by the solid curve.  $E_{\rm rad} $ changes from zero to a constant value
$\simeq 0.00925/8\pi G_5\omega_0^2$ as $r_{\rm obs} \omega_0$ increases from
zero to 40.  The value of $E_{\rm rad}$ near the center 
$r_{\rm obs} \omega_0 \sim 1$ does not have a definite meaning because the
Landau-Lifshitz pseudotensor is not gauge invariant.  However, the
asymptotic value of $E_{\rm rad}$ for $r_{\rm obs} \omega_0 \gg 1$ is gauge
invariant and should indicate the correct amount of the radiated energy.  
Note that because of the conservation law
\eqref{conservation-perturbation}, $\partial_\mu t_{\rm
LL}^{\mu\nu}=0$, the integrated energy flux $E_{\rm rad}(r_{\rm obs})$
has to be equal to the initial amount of energy within the surface
$r=r_{\rm obs}$, 
%
\begin{equation} 
E(r_{\rm obs})=\int_{r\le r_{\rm obs}} t_{\rm LL}^{00}dV 
\end{equation} 
%
where $dV$ is a volume element of the 4D space.  This
is directly checked by the numerical integrations.

In order to calculate $E_{\rm rad}(r_{\rm obs})$ for the numerical solution, 
we proceed as follows:  At each time step, we define the 
perturbed quantities as $h_{tt}=-2(\alpha-1)$, $h_{ti}=\beta_i$, and
$h_{ij}=\gamma_{ij}-\delta_{ij}$, and evaluate $t^{i0}$ at $r=r_{\rm
  obs}$ using Eq.~\eqref{tLL-perturbation}.  Then, we calculate the integral 
%
\begin{equation}
\frac{dE_{\rm rad}}{dt}=\int t^{i0} \hat n_idS, 
\end{equation}
%
that gives the energy flux through a surface $r=r_{\rm obs}$, where
$\hat n_i$ is the outward unit normal to the surface.  Finally,
$dE_{\rm rad}/dt$ is integrated from the initial time to the final
time to obtain $E_{\rm rad}(r_{\rm obs})$. 
In order to evaluate the metric functions
for a surface of $r=r_{\rm obs}$, we used linear,
quadratic and cubic interpolations and compared the results.
Although relatively large deviation from 
the analytic result is seen for the linear interpolation, 
the deviation becomes smaller
when the quadratic interpolation is used.
The result of the cubic interpolation did not improve
that of the quadratic interpolation. This is because in these cases,
the error primarily comes from the error generated at the outer boundary
(i.e., the error due to the inaccuracy of the outgoing boundary condition)
which is approximately at the second order with respect to the grid size.

Figure~\ref{RE_linear_gaussian_onepluslog} shows the results of $8\pi
G_5 E_{\rm rad} \omega_0^2$ evaluated by the quadratic
interpolation at several observation points 
for the grid resolutions
$\Delta x=0.1$ ($\odot$), $0.15$ ($\blacktriangle$), and $0.2$ ($\times$).
The deviation decreases as the resolution is increased 
and the numerical data approach the analytic result. 

We also evolved the gravitational-wave packet using the dynamical
slicing and $\Gamma$-driver conditions and checked that $E_{\rm rad}$
depends very weakly on the initial choice of the lapse and shift as
long as the initial value of $\alpha-1$ and $\beta^i$ is small.  This
is natural because the gauge invariance of $E_{\rm rad}$ is guaranteed
for $r_{\rm obs} \omega_0 \gg 1$.  Therefore, we conclude that the extraction of
gravitational-wave energy flux by the Landau-Lifshitz pseudotensor
works well, as far as the amplitude of gravitational waves is
sufficiently small at the extracted region (i.e., in the wave zone).

%
%
\section{Summary}

This paper describes the formulations for numerical relativity in
higher dimensions and reports the new codes for simulating 5D
spacetimes.  We derived the BSSN formalism for higher-dimensional
spacetimes  and
also studied the cartoon method in 5D spacetimes of $U(1)$ symmetry
($x=y,z,w$), $U(1)\times U(1)$ symmetry ($x=y$, $z=w$), and
$SO(3)$ symmetry ($x=y=z$, $w$).  Based on the BSSN formalism and
the cartoon methods, we have implemented the new 5D numerical relativity
codes, and tested them by evolving the 5D Schwarzschild spacetime and the
spacetime composed of gravitational waves of small amplitude.  The
numerical results converge to the analytic solutions with improving
the grid resolution at the correct order (fourth order).  It was also
demonstrated that the 5D Schwarzschild spacetime can be evolved for a
long time by the puncture approach, as in the 4D case.

We proposed the method of extracting gravitational-wave energy flux by
the Landau-Lifshitz pseudotensor and tested this method. We showed that
the integrated energy fluxes calculated at several surfaces $r=r_{\rm
obs}$ agree well with the semianalytic solution derived by perturbative
calculations. Furthermore, it was confirmed that the result is
insensitive to the gauge conditions for the lapse and shift. These
results indicate that the energy extraction by the Landau-Lifshitz
pseudotensor works well.  The remaining problem would be to check
that the extraction in the wave zone works well even when the central
region is highly nonlinear, e.g., a head-on collision of two black
holes. This will be tested by performing a simulation of Brill wave
spacetime in five dimensions and by comparing the ADM mass of the initial
data and the energy radiated during the evolution.  Also, it is
necessary to check if the extraction of the angular momentum is
possible.  We expect that the radiated angular momentum also can be
calculated by the Landau-Lifshitz pseudotensor using a similar manner
to the 4D case \cite{T80}.

As discussed in Sec. I, there are many interesting issues of nonlinear
dynamics in higher-dimensional gravity, which should be studied in
numerical relativity. In this paper, we have prepared the tools
necessary for simulating higher-dimensional spacetimes.  Our next step
is to apply our codes to the unsolved problems. 

\acknowledgments

We thank K.~Nakao, T.~Shiromizu, and T.~Tanaka for discussions,
and HY also thanks V.~Cardoso for comments.
Numerical computations were in part performed on the NEC-SX9 at CfCA
in National Astronomical Observatory of Japan and on the NEC-SX8 at
Yukawa Institute for Theoretical Physics (YITP) at Kyoto 
University. HY thanks YITP, where this work was
initiated, for financial support and hospitality.  
MS was in part supported by Grant-in-Aid for Scientific Research 
(21340051) and by Grant-in-Aid for Scientific Research on Innovative
Area (20105004) of the Japanese Monbukagakusho.

\appendix

%
%
\section{Linear gravitational waves in the 5D spacetime}

In this section, we describe solutions of the linearized Einstein
equations in the 5D flat spacetime focusing on the perturbations
preserving $U(1)\times U(1)$ symmetry and $SO(3)$ symmetry.  In the
following, we denote the metric perturbation as $h_{\mu\nu}$, which
obeys the linearized Einstein equation
%
\begin{equation}
\delta G_{\mu\nu}[h_{\alpha\beta}]=0. \label{linearG}
\end{equation}
%

The analysis for perturbations of higher-dimensional Schwarzschild
black holes described in Ref.~\cite{Kodama:2003jz} is partially used,
since this formulation is applicable also for the flat spacetime. In
their approach, the perturbation is decomposed into the scalar,
vector, and tensor modes (with respect to the 3D unit sphere) using
spherical harmonic functions, and the master equations are 
derived for the gauge-invariant variables. We adopt their method for 
spherical harmonic expansion but do not use the master equations,
because we are interested in explicit special solutions for which 
the master equations are not necessary. 

%
%
\subsection{Perturbation with $U(1)\times U(1)$ symmetry}

We here derive a solution of $U(1)\times U(1)$ symmetry for 
Eq.~\eqref{linearG}.  Denoting the coordinates by $(t, x, y, z, w)$,
we introduce the following curvilinear coordinates:
%
\begin{eqnarray}
x &=& r \sin\theta \cos\varphi, \\
y &=& r \sin\theta \sin\varphi, \\
z &=& r \cos\theta \cos\psi, \\
w &=& r \cos\theta \sin\psi.
\end{eqnarray}
%
In these coordinates, the line element of the flat space is 
%
\begin{eqnarray}
dl^2=dr^2 + r^2 (d\theta^2 + \sin^2\theta d \varphi^2
+\cos^2\theta d \psi^2 ). 
\end{eqnarray}
%
For the $U(1)\times U(1)$ symmetry case, 
the linear perturbation $h_{\mu\nu}$ satisfies 
%
\begin{eqnarray}
\frac{\partial h_{\mu\nu}}{\partial \varphi}=
\frac{\partial h_{\mu\nu}}{\partial \psi}=0, \label{eq00}
\end{eqnarray}
%
since $\partial_{\varphi}$ and $\partial_{\psi}$ are the
Killing vectors.

In the gauge condition with $\alpha=1$ and $\beta^k=0$ (i.e.,
$h_{00}=h_{0i}=0$), the spatial components of the linear perturbation
$h_{ij}$ satisfy
%
\begin{eqnarray}
\ddot h_{ij}=\Delta h_{ij}, \label{eq01}
\end{eqnarray}
%
where $\Delta$ is the flat 4D Laplacian. The Hamiltonian 
and momentum constraints in the linear approximation give 
%
\begin{eqnarray}
&& \bar{g}^{ij} h_{ij}=0, \label{eq02} \\
&& \bar{D}^i h_{ij}=a_j, \label{eq03}
\end{eqnarray}
%
where $\bar{g}_{ij}$ is the 4D flat space metric
in the curvilinear coordinates and 
$\bar{D}_i$ is the covariant derivative with respect to $\bar{g}_{ij}$. 
$a_j$ denotes a constant vector determined at the initial state, 
which is set to be zero in the following for simplicity.

%
%
\subsubsection{Scalar mode}

The scalar mode (with respect to a 3D unit sphere) is expanded in 
terms of scalar harmonic functions $\mathbb{S}$ on a 3D unit sphere 
with the metric $d\sigma^2=\sigma_{IJ}dz^Idz^J=d\theta^2 +\sin^2\theta
d\varphi^2+\cos^2\theta d\psi^2$, which satisfy the equation
%
\begin{equation}
[\hat{\Delta}_3+l(l+2)]\mathbb{S}=0,
\label{scalar-harmonic-eq}
\end{equation}
%
where $\hat{\Delta}_3=\hat{D}^I\hat{D}_I$ is the Laplace operator on
the 3D unit sphere.  In the following, we focus only on solutions of
the lowest-order multipole moment $l=2$, for which the harmonic
function is
%
\begin{equation}
\mathbb{S}=2\cos^2\theta-1.
\end{equation}
%
The scalar-mode perturbation is
given in the form 
%
\begin{equation}
h_{ij}=\left(
\begin{array}{cc}
a(t,r)\mathbb{S} & rb(t,r)\mathbb{S}_{J} \\
* & r^2\left[c(t,r) \mathbb{S}\sigma_{IJ} + d(t,r)\mathbb{S}_{IJ}  \right] 
\end{array}
\right),
\label{metric-scalar-mode}
\end{equation} 
%
where 
%
\begin{equation}
\mathbb{S}_J:=\hat{D}_J\mathbb{S};
\qquad \textrm{and} \qquad 
\mathbb{S}_{IJ}:=\frac{1}{l(l+2)}\hat{D}_I\hat{D}_J\mathbb{S}
+\frac{1}{3}\sigma_{IJ}\mathbb{S},
\label{def-Si-Sij}
\end{equation} 
%
or more explicitly,
%
\begin{equation}
\mathbb{S}_J=(-4\sin\theta\cos\theta,\ 0,\ 0),
\end{equation}
%
\begin{equation}
\mathbb{S}_{IJ}=\frac{1}{6} \times
\left(
\begin{array}{ccc}
1-2\cos^2\theta & 0 & 0 \\
* & \sin^2\theta(\cos^2\theta-2) & 0 \\
* & * & \cos^2\theta(1+\cos^2\theta)
\end{array}
\right).
\end{equation}
%
Equation~\eqref{eq02} gives $a+3c=0$, and Eq.~\eqref{eq03} yields 
%
\begin{equation}
8b=4a+ra_{,r}, 
\label{b-and-a}
\end{equation}
%
%
\begin{equation}
\frac{5}{12}d=4b+rb_{,r}+c.
\label{d-and-b-c}
\end{equation}
%
The $rr$ component of Eq.~\eqref{eq01} with Eq.~\eqref{b-and-a} 
gives a wave equation for $a$:
%
\begin{equation}
\ddot{a}=a_{,rr}+\frac{7}{r}a_{,r}.
\label{eq:a}
\end{equation}
%
Equations~\eqref{b-and-a} and \eqref{d-and-b-c} imply that
once $a$ is computed from the equation \eqref{eq:a}, 
$b$, $c$ and $d$ are subsequently determined.
The obtained solution is guaranteed to satisfy 
other components of Eq.~\eqref{eq01}.

Defining $a \equiv u/r^3$, we obtain the equation 
%
\begin{eqnarray}
\ddot u =u_{,rr} + \frac{1}{r} u_{,r}-\frac{n^2}{r^2} u.
\label{eq-for-u}
\end{eqnarray}
%
Here, $n=l+1=3$. The formal solution of this equation 
is written as
%
\begin{eqnarray}
u={\rm Re}\left[\int d\omega f(\omega) e^{i \omega t} J_n(\omega r)\right],
\label{equ}
\end{eqnarray}
%
where $f(\omega)$ is an arbitrary function of $\omega$, and $J_n$ is 
the Bessel function of $n$-th order. In the integral expression, 
it is written by 
%
\begin{equation}
J_n(z)=\frac{1}{2\pi} 
\int_0^{2\pi} d\vartheta \cos(n\vartheta-z\sin\vartheta). 
\end{equation}
%
To constitute a solution for the propagation of a gravitational-wave packet, 
we set $f(\omega)=-i \sqrt{2\pi} A_0 e^{-\omega^2/2\omega_0^2}$. 
Then, Eq.~\eqref{equ} is integrated to give
%
\begin{eqnarray}
u(t,r)=A_0 \omega_0 
\int^{2\pi}_0 d\vartheta \sin(n\vartheta)e^{-\omega_0^2(t-r\sin\vartheta)^2/2}.
\label{special-solution-u} 
\end{eqnarray}
%
In this case, $u=0$ at $t=0$, and thus $h_{ij}=0$, whereas 
the extrinsic curvature $K_{ij}=-\dot h_{ij}/2$ is not zero because 
%
\begin{eqnarray}
\dot u(0,r)=r A_0 \omega_0^3 
\int^{2\pi}_0 d\vartheta \sin(n\vartheta)\sin\vartheta
e^{-(\omega_0 r\sin\vartheta)^2/2}
\end{eqnarray}
%
is not zero at $t=0$.

%
%
\subsubsection{Vector mode}

Perturbation $h_{ij}$ of the vector type is expanded 
in terms of the harmonic vectors
$\mathbb{V}_{I}$ satisfying
%
\begin{equation}
[\hat{\Delta}_3+l(l+2)-1]\mathbb{V}_{I}=0,
\end{equation}
%
%
\begin{equation}
\hat{D}_J\mathbb{V}^{J}=0.
\end{equation}
%
Under the $U(1)\times U(1)$ symmetry, only the modes for odd $l$ numbers 
are nonzero. Since the $l=1$ mode denotes a stationary perturbation with
angular momentum, the lowest value of $l$ is 3 for the nonstationary perturbation. 
For this mode,
%
\begin{equation}
\mathbb{V}^J=\left(0, \ A(\sin^2\theta-2/3), \ B(\cos^2\theta-2/3)\right),
\end{equation}
%
where $A$ and $B$ are arbitrary constants. 
The perturbation is given in the form
%
\begin{equation}
h_{ij}=\left(
\begin{array}{cc}
\ 0 \ & (1/r)k(t,r) \mathbb{V}_J \\
* & r o(t,r)\mathbb{V}_{IJ} 
\end{array}
\right),
\label{metric-vector-mode}
\end{equation} 
%
where $\mathbb{V}_{IJ}$ is defined by
%
\begin{eqnarray}
\mathbb{V}_{IJ}&:=&\frac12
\left(\hat{D}_{I}\mathbb{V}_{J}+\hat{D}_{J}\mathbb{V}_{I}\right) 
\nonumber \\
&=&\left(
\begin{array}{ccc}
\ 0 & \ A \sin^3\theta\cos\theta & \ -B\sin\theta\cos^3\theta\\
\ * & 0 & 0 \\
\ * & * & 0 
\end{array}
\right).
\end{eqnarray}
%
From Eqs.~\eqref{eq01} and \eqref{eq03}, 
the equations for $k$ and $o$ are derived as 
%
\begin{equation}
\ddot{k}=k_{,rr}+\frac{1}{r}k_{,r}-\frac{16}{r^2}k,
\label{eq-for-k}
\end{equation}
%
\begin{equation}
6o =  k_{,r} + \frac{2}{r}k.
\label{eq-for-i}
\end{equation}
%
Here, Eq.~\eqref{eq-for-k} has the same form as Eq.~\eqref{eq-for-u}
but with $n=4$. Hence, a special solution for $k(t,r)$ is given by
Eq.~\eqref{special-solution-u} with $n=4$, and then 
$o(t,r)$ is calculated from Eq.~\eqref{eq-for-i}.

%
%
\subsubsection{Tensor mode}

Perturbation $h_{ij}$ of the tensor type is expanded 
in terms of the harmonic tensors
$\mathbb{T}_{IJ}$ satisfying
%
\begin{equation}
[\hat{\Delta}_3+l(l+2)-2]\mathbb{T}_{IJ}=0,
\end{equation}
%
%
\begin{equation}
\mathbb{T}^{I}_{~I}=0,\qquad
\hat{D}_J\mathbb{T}^{J}_{~I}=0.
\end{equation}
%
Under $U(1)\times U(1)$ symmetry, the possible harmonic
tensors for $l=2$ are
%
\begin{equation}
\mathbb{T}_{IJ}=
\left(
\begin{array}{ccc}
\ A \ & 0 & 0 \\
* & A\sin^2\theta(1-3\sin^2\theta) & B\sin^2\theta\cos^2\theta \\
* & * & A\cos^2\theta(3\sin^2\theta-2)
\end{array}
\right),
\label{harmonic-tensor}
\end{equation}
%
where $A$ and $B$ are arbitrary constants.
The perturbation is given in the form
%
\begin{equation}
h_{ij}=\left(
\begin{array}{cc}
\ 0 \ &  \ 0 \ \\
* & rh(t,r)\mathbb{T}_{IJ} 
\end{array}
\right),
\label{metric-tensor-mode}
\end{equation} 
%
and from Eq.~\eqref{eq01}, the equation for $h$ becomes
%
\begin{equation}
\ddot{h}=h_{,rr}+\frac{1}{r}h_{,r}-\frac{9}{r^2}h.
\end{equation}
%
This is the same equation as Eq.~\eqref{eq-for-u}, and thus
a special solution is given by Eq.~\eqref{special-solution-u}
with $n=3$.

%
%
\subsection{Perturbation with $SO(3)$ symmetry}

Next, we derive solutions of a perturbation of $SO(3)$ symmetry. 
For the Minkowski coordinates $(t, x, y, z, w)$, 
we introduce the following hyperspherical coordinates: 
%
\begin{eqnarray}
x &=& r \sin\theta \sin\varphi \sin\psi , \\
y &=& r \sin\theta \sin\varphi \cos\psi , \\
z &=& r \sin\theta \cos\varphi , \\
w &=& r \cos\theta .
\end{eqnarray}
%
Then, the line element of the flat space is given by
%
\begin{eqnarray}
dl^2=dr^2 + r^2 (d\theta^2 + \sin^2\theta d \varphi^2
+\sin^2\theta\sin^2\varphi d \psi^2 ). 
\end{eqnarray}
%
Here, we consider solutions of $SO(3)$ 
symmetry with the Killing vectors 
%
\begin{eqnarray}
\boldsymbol{\xi}_1&=&
-\cos\psi\partial_\varphi+\cot\varphi\sin\psi\partial_\psi,\\
\boldsymbol{\xi}_2&=&
\sin\psi\partial_\varphi+\cot\varphi\cos\psi\partial_\psi,\\
\boldsymbol{\xi}_3&=&
\partial_\psi. 
\end{eqnarray}
%
Under the requirement of this symmetry, the vector and tensor modes do
not exist, because there are no vector and tensor harmonic functions that
satisfy $\mathcal{L}_{\xi_n}\mathbb{V}_{I}=0$ and 
$\mathcal{L}_{\xi_n}\mathbb{T}_{IJ}=0$.  Therefore, only the scalar
mode should be analyzed. 

The scalar harmonic function defined by Eq.~\eqref{scalar-harmonic-eq}
on a 3D unit sphere with the metric
$d\sigma^2=\sigma_{IJ}dz^Idz^J=d\theta^2+\sin^2\theta d\varphi^2
+\sin^2\theta \sin^2\varphi d\psi^2$ is  
%
\begin{equation}
\mathbb{S}=4\cos^2\theta-1, \quad \textrm{for} \quad l=2. 
\end{equation}
%
The metric perturbation is given in the same form as
Eq.~\eqref{metric-scalar-mode}. Here, definitions for $\mathbb{S}_I$
and $\mathbb{S}_{IJ}$ are same as Eq.~\eqref{def-Si-Sij}, and
their explicit forms in this case are
%
\begin{equation}
\mathbb{S}_J=(-8\sin\theta\cos\theta, \ 0, \ 0),
\end{equation}
%
%
\begin{equation}
\mathbb{S}_{IJ}=\frac{1}{3}\sin^2\theta \times
\left(
\begin{array}{ccc}
2 & 0 & 0 \\
* & -\sin^2\theta & 0 \\
* & * & -\sin^2\theta\sin^2\varphi 
\end{array}
\right).
\end{equation}
%
The equations for $a$, $b$, $c$, and $d$ are the same as
Eqs.~\eqref{b-and-a}--\eqref{eq:a}, and a special solution for
$u=r^3a$ is given by the same formula as
Eq.~\eqref{special-solution-u}.

%
%
\section{Landau-Lifshitz pseudotensor}

In this section, we derive the Landau-Lifshitz pseudotensor in a
$D$-dimensional spacetime ${\cal M}$ with the metric $g_{\mu\nu}$.
Following \cite{LL75}, we define
%
\begin{equation}
\tilde{g}^{\mu\nu}=\sqrt{-g}g^{\mu\nu},
\label{def-tilde-g}
\end{equation}
%
where $g$ is the determinant of the metric, and then we introduce the 
super-potential
%
\begin{equation}
H^{\mu\alpha\nu\beta}=
\tilde{g}^{\mu\nu}\tilde{g}^{\alpha\beta}
-\tilde{g}^{\alpha\nu}\tilde{g}^{\mu\beta}.
\end{equation}
%
The Landau-Lifshitz pseudotensor is defined by
%
\begin{equation}
16\pi G_Dt^{\mu\nu}_{\rm LL}
=(-g)^{-1}H^{\mu\alpha\nu\beta}_{~~~~~,\alpha\beta}
-\left(2R^{\mu\nu}-g^{\mu\nu}R\right).
\end{equation}
%
From this definition, the conservation law is derived:
%
\begin{equation}
\left[(-g)\left(T^{\mu\nu}+t^{\mu\nu}_{\rm LL}\right)\right]_{,\nu}=0.
\label{conservation-law}
\end{equation}
%
Because the Landau-Lifshitz pseudotensor is not a tensor, it does not
have a covariant meaning in general. However, for a perturbed flat
spacetime, the leading-order terms of $t_{\rm LL}^{\mu\nu}$ with
respect to the perturbative quantities can be used to evaluate the
total energy and total radiated energy of the gravitational field in a
gauge-invariant manner (see below).

In Ref.~\cite{LL75}, two expressions for $t_{\rm LL}^{\mu\nu}$
are given. The first one is the expression in terms
of the Christoffel symbols, and this expression holds
for arbitrary dimensionality $D$. The second one is the 
expression by the metric functions, and it is modified to give
%
\begin{multline}
16\pi G_D(-g)t_{\rm LL}^{\mu\nu}=
\tilde{g}^{\mu\nu}_{~~,\alpha}\tilde{g}^{\alpha\beta}_{~~,\beta}
-\tilde{g}^{\mu\alpha}_{~~,\alpha}\tilde{g}^{\nu\beta}_{~~,\beta}+
\frac12g^{\mu\nu}g_{\alpha\beta}
\tilde{g}^{\alpha\rho}_{~~,\sigma}\tilde{g}^{\sigma\beta}_{~~,\rho}
\\
-\left(
g^{\mu\alpha}g_{\beta\rho}\tilde{g}^{\nu\rho}_{~~,\sigma}\tilde{g}^{\beta\sigma}_{~~,\alpha}
+g^{\nu\alpha}g_{\beta\rho}\tilde{g}^{\mu\rho}_{~~,\sigma}\tilde{g}^{\beta\sigma}_{~~,\alpha}
\right)
+g_{\alpha\beta}g^{\rho\sigma}
\tilde{g}^{\mu\alpha}_{~~,\rho}\tilde{g}^{\nu\beta}_{~~,\sigma}
\\
+\frac{1}{4(D-2)}
\left(2g^{\mu\alpha} g^{\nu\beta}-g^{\mu\nu}g^{\alpha\beta}\right)
\left[(D-2)g_{\rho\sigma}g_{\gamma\delta}-g_{\sigma\gamma}g_{\rho\delta}\right]
\tilde{g}^{\rho\delta}_{~~,\alpha}\tilde{g}^{\sigma\gamma}_{~~,\beta}
\label{tLL-metric}
\end{multline}
%
in $D$ dimensions.
Let us consider the perturbation on a flat spacetime,
whose metric is $g_{\mu\nu}=\eta_{\mu\nu}+h_{\mu\nu}$,
where $\eta_{\mu\nu}$ is the flat metric in the Minkowski
coordinates. Defining 
$\hat{h}_{\mu\nu}:=h_{\mu\nu}-(1/2)h\eta_{\mu\nu}$,
the Landau-Lifshitz pseudotensor is rewritten as
%
\begin{multline}
16\pi G_Dt_{\rm LL}^{\mu\nu}=
\hat{h}^{\mu\nu}_{~~,\alpha}\hat{h}^{\alpha\beta}_{~~,\beta}
-\hat{h}^{\mu\alpha}_{~~,\alpha}\hat{h}^{\nu\beta}_{~~,\beta}
+
\frac12\eta^{\mu\nu}
\hat{h}^{\alpha\rho}_{~~,\sigma}\hat{h}^{\sigma}_{~\alpha,\rho}
\\
-\left(
\hat{h}^{\mu\rho}_{~~,\sigma}\hat{h}_\rho^{~\sigma,\nu}
+\hat{h}^{\nu\rho}_{~~,\sigma}\hat{h}_\rho^{~\sigma,\mu}
\right)
+
\hat{h}^{\mu\alpha,\rho}\hat{h}^{\nu}_{~\alpha,\rho}
\\
+\frac12\hat{h}^{\rho\sigma,\mu}\hat{h}_{\rho\sigma}^{~~,\nu}
-\frac14\eta^{\mu\nu}\hat{h}^{\rho\sigma,\alpha}\hat{h}_{\rho\sigma,\alpha}
-\frac{1}{4(D-2)}\left(2\hat{h}^{,\mu}\hat{h}^{,\nu}-\eta^{\mu\nu}\hat{h}^{,\alpha}\hat{h}_{,\alpha}\right).
\label{tLL-perturbation}
\end{multline} 
%
Here, we have kept only the second-order quantities 
of the perturbation. 
Note that the second-order Landau-Lifshitz pseudotensor $t^{\mu\nu}_{\rm LL}$ 
behaves as a tensor against the general coordinate
transformations of the background spacetime (but for a fixed gauge),
by replacing the coordinate
derivatives $(,\mu)$ to the covariant derivatives
and the Minkowski metric $\eta^{\mu\nu}$ to the flat background metric
$\bar{g}^{\mu\nu}$ in curved coordinates
in Eq.~\eqref{tLL-perturbation}.
The quantity $t^{0r}$ in Eq.~\eqref{Erad-linear}
has to be evaluated in this way.

For the expression \eqref{tLL-perturbation}, the conservation
law~\eqref{conservation-law} for a vacuum spacetime
becomes 
%
\begin{equation}
\partial_\mu t_{\rm LL}^{\mu\nu}=0
\label{conservation-perturbation}
\end{equation}
%
in the Minkowski coordinates, which suggests that $t_{\rm LL}^{\mu\nu}$ can
be interpreted as the effective stress-energy tensor of the
gravitational field valid up to second order in $h_{\mu\nu}$. 
Here, we have to be careful because the Landau-Lifshitz
pseudotensor is not the unique quantity satisfying the condition
\eqref{conservation-perturbation} and also because this quantity is not
gauge invariant (see Ref.~\cite{Wald}).  However, the total energy
%
\begin{equation}
E_{\rm tot}=\int t_{\rm LL}^{00}dV
\label{Etot}
\end{equation}
%
is shown to be the gauge-invariant quantity,
where $dV$ is the volume element of the
spacelike hypersurface. Similarly, the total radiated energy
%
\begin{equation}
E_{\rm rad}=\int t^{0i}\hat{n}_idSdt
\label{Erad}
\end{equation}
%
is gauge-invariant, where $dS$ and $\hat{n}^i$ are the area element
and an outward unit normal of a surface at the distant region.
Therefore, the Landau-Lifshitz pseudotensor $t_{\rm LL}^{\mu\nu}$
provides us a reliable method for evaluating the total radiated energy.



\end{document}